%% file: main.tex
\documentclass[10pt, conference]{IEEEtran}

\usepackage[utf8]{inputenc}
\usepackage{color}
\usepackage{graphicx}
\usepackage{listings}
\usepackage{etoolbox}
\usepackage{tabularx}
\usepackage{booktabs}
\usepackage[caption=false]{subfig}
\usepackage[export]{adjustbox}
\usepackage{balance}

\usepackage{amsmath}
\usepackage{amssymb}
\usepackage{mathtools}
\usepackage[ruled,shortend,linesnumbered]{algorithm2e}

\usepackage{rotating}
\usepackage{multirow}
\usepackage{pifont}

\usepackage{tikz}
\usetikzlibrary{calc, arrows, automata, patterns, shadings, shadows, shapes.arrows, positioning}

\definecolor{purpleboxcolor}{RGB}{230,224,236}
\definecolor{greenboxcolor}{RGB}{235,241,222}
\definecolor{beigeboxcolor}{RGB}{221,217,195}
\definecolor{whiteboxcolor}{RGB}{255,255,255}
\definecolor{blueboxcolor}{RGB}{219,238,244}
\definecolor{orangeboxcolor}{RGB}{252,213,181}
\definecolor{defcolor}{RGB}{255,228,179}
\definecolor{controllerblue}{RGB}{198,217,241}
\definecolor{buildingblockgray}{RGB}{242,242,242}
\definecolor{commentblue}{RGB}{9,134,176}
\definecolor{delegationred}{RGB}{242,220,219}

\newcommand*\circled[1]{\raisebox{.5pt}{\textcircled{\raisebox{-.9pt} {#1}}}}

\providetoggle{cmts}
\settoggle{cmts}{true} 

\IEEEoverridecommandlockouts
\begin{document}

\title{An Optimization-based Approach for Flow Table Capacity Bottleneck Mitigation in Software-Defined Networks}

\author{
  \IEEEauthorblockN{
    Robert Bauer\IEEEauthorrefmark{1},
    Martina Zitterbart\IEEEauthorrefmark{1}
  }
  \IEEEauthorblockA{
    \IEEEauthorrefmark{1}Karlsruhe Institute of Technology, Germany\\
    Email: \{robert.bauer$\mid$zitterbart\}\symbol{64}kit.edu
  }
}

\maketitle

\begin{abstract}
Flow delegation is a flexible technique to mitigate flow table capacity bottlenecks in Software-defined Networks (SDN).
Such bottlenecks occur when SDN switches provide insufficient flow table capacity which leads to performance degradation and network failures. Flow delegation addresses this problem by automatically relocating flow rules from a bottlenecked switch to neighboring switches with spare capacity.
This paper introduces a new algorithm to efficiently perform flow delegation based on a novel delegation template abstraction and multi-period multi-objective optimization.
Different from existing work, our approach can include estimated knowledge about future network situations and deal with different optimization criteria such as link and control overhead. 
We discuss the problem decomposition for the new algorithm and introduce an efficient two-step heuristic.
Results show, that our approach performs significantly better than the simple greedy algorithm used in earlier work and is capable of handling flow delegation for networks with hundreds of switches. 
\end{abstract}


\IEEEpeerreviewmaketitle

\input{macros} 

\input{sections/introduction}

\input{sections/example}
\input{sections/functionality}

\input{sections/evaluation}

\input{sections/related_work}

\input{sections/conclusion}

\bibliographystyle{IEEEtran}
\bibliography{lit2}

\end{document}

%% file: macros.tex
\newcommand\ddfrac[2]{\frac{\displaystyle #1}{\displaystyle #2}}

\newcommand{\FigurePartTwo}[4]{
  \begin{figure}[#1]
    \centering
    \includegraphics[width=#2\columnwidth]{part2/figures/#3}
    \caption{#4.}\label{fig:part2:#3}
  \end{figure}
}

\newcommand{\FigureEval}[4]{
  \begin{figure}[#1]
    \centering
    \includegraphics[width=#2\columnwidth]{part2/figures/eval/#3}
    \caption{#4.}\label{fig:#3}
  \end{figure}
}

\newcommand{\STAB}[1]{\begin{tabular}{@{}c@{}}#1\end{tabular}}

\newcommand{\tauInstall}{
\begin{tikzpicture}
     \draw node[preaction={fill, yellow},draw=black,line width=0.5mm,circle,minimum size=0.15cm,inner sep=3pt] (1) at(0,0) {};
\end{tikzpicture}
}

\newcommand{\tauRemove}{
\begin{tikzpicture}
     \draw node[preaction={fill, yellow},draw=black,line width=0.5mm,pattern=north east lines, pattern color=black,circle,minimum size=0.15cm,inner sep=3pt] {};
\end{tikzpicture}
}

\newcommand{\ruleBoxAgg}{
\begin{tikzpicture}
     \draw node[preaction={fill, black},draw=black,line width=0.5mm,pattern=north east lines, pattern color=black,rectangle,minimum size=0.5cm,minimum height = 0.3cm,inner sep=3pt] {};
\end{tikzpicture}
}

\newcommand{\ruleBoxBackflow}{
\begin{tikzpicture}
     \draw node[preaction={fill, white},draw=black,line width=0.5mm,pattern=north west lines, pattern color=black,rectangle,minimum size=0.5cm, minimum height = 0.3cm, inner sep=3pt] {};
\end{tikzpicture}
}

\newcommand{\ruleBoxRelocated}{
\begin{tikzpicture}
     \draw node[preaction={fill, yellow},draw=black,line width=0.5mm,pattern=north east lines, pattern color=black,rectangle,minimum size=0.5cm,minimum height = 0.3cm,inner sep=3pt] {};
\end{tikzpicture}
}

\newcommand{\ruleBoxRegular}{
\begin{tikzpicture}
     \draw node[preaction={fill, yellow},draw=black,line width=0.5mm,rectangle,minimum size=0.5cm,minimum height = 0.3cm,inner sep=3pt] {};
\end{tikzpicture}
}

\makeatletter
\newcommand{\removelatexerror}{\let\@latex@error\@gobble}
\makeatother

\makeatletter
\newcommand\ztag[1]{%
\def\@currentlabel{#1}%
\gdef\tmp{%
\addtocounter{equation}{-1}%
\def\theequation{#1}}%
\aftergroup\aftergroup\aftergroup\aftergroup\aftergroup\aftergroup
\aftergroup\aftergroup\aftergroup\aftergroup\aftergroup\aftergroup
\aftergroup\aftergroup\aftergroup\aftergroup\aftergroup\aftergroup
\aftergroup\aftergroup\aftergroup\aftergroup\aftergroup\aftergroup
\aftergroup\aftergroup\aftergroup\aftergroup\aftergroup\aftergroup
\aftergroup
\tmp}
\makeatother

\catcode`@=11
\def\caseswithdelim#1#2{\left#1\,\vcenter{\normalbaselines\m@th
  \ialign{\strut$##\hfil$&\quad##\hfil\crcr#2\crcr}}\right.}
\catcode`@=12
\def\bcases#1{\caseswithdelim[{#1}}
\def\vcases#1{\caseswithdelim|{#1}}

\newcommand*\rectangled[1]{%
\tikz[baseline=(R.base)]\node[draw,rectangle,inner sep=2pt](R) {#1};\!
}

\newcommand*\rectangledDef[1]{%
\tikz[baseline=(R.base)]\node[draw,rectangle,fill=lightgray!40,inner sep=2pt](R) {#1};\!
}

\newcommand*\rectangledC[1]{%
\tikz[baseline=(R.base)]\node[draw,rectangle,fill=white,inner sep=3pt](R) {#1};\!
}

\newcommand{\evalpath}{/home/bauer/Code/PipeFlex/plugins/de.bauer.research.delegation/standalone/eval/plots}

\newcommand{\vvv}{\vspace{0.8em}}

\newcommand{\rasf}{\mathit{ASF}}
\newcommand{\asf}[1][\np]{\rasf_{#1}}
\newcommand{\asfc}[1][\np]{\asf[#1]^c}
\newcommand{\asfe}[1][\np]{\asf[#1]^e}

\newcommand{\pluseq}{\mathrel{{+}{=}}}
\newcommand{\minuseq}{\mathrel{{-}{=}}}

\newcommand{\objCapacity}{\texttt{O1-CAPACITY}}
\newcommand{\objFeature}{\texttt{O2-FEATURE}}
\newcommand{\objControl}{\texttt{O3-CONTROL}}

\SetKwProg{Algo}{Algorithm}{}{}
\SetKwFunction{Assert}{assert}
\SetKwFor{Loop}{loop}{}{end}
\SetKw{Continue}{continue}
\SetKw{Break}{break}
\SetKw{Yield}{yield}
\SetKw{True}{true}
\SetKw{False}{false}
\SetKw{Not}{not}
\SetKw{And}{and}
\SetKw{Or}{or}

\newcommand{\makespace}{\vspace*{0.7em}}

\newcommand{\citePBCE}{\cite{own-pbce}} 

\newcommand{\setTo}{\ensuremath{\leftarrow\ }} 

\newcommand{\hdst}{h_\texttt{dst}}
\newcommand{\hsrc}{h_\texttt{src}}
\newcommand{\macsrc}{\texttt{mac\_src}}
\newcommand{\macdst}{\texttt{mac\_dst}}
\newcommand{\fwd}{\texttt{fwd}}
\newcommand{\ttindex}{\texttt{index}}
\newcommand{\set}{\texttt{set}}
\newcommand{\CS}{\texttt{CS}}

\newcommand{\flagrr}{\texttt{remote\_rule\_indicator}}
\newcommand{\flagbf}{\texttt{backflow\_indicator}}
\newcommand{\flagip}{\texttt{inport\_indicator}}
\newcommand{\flagrrs}{\texttt{rri}}

\newcommand{\rri}{\rectangledC{\texttt{rri}}}
\newcommand{\bfi}{\rectangledC{\texttt{bfi}}}
\newcommand{\ipi}{\rectangledC{\texttt{ipi}}}

\newcommand{\Rri}{\texttt{rri}}
\newcommand{\Bfi}{\texttt{bfi}}
\newcommand{\Ipi}{\texttt{ipi}}

\newcommand{\calculateDSS}{\textsc{calculate\_delegation\_status}~}

\newcommand{\ODT}{\ensuremath{\texttt{Obj}_\texttt{DT}}}
\newcommand{\ORI}{\ensuremath{\texttt{Obj}_\texttt{RI}}}
\newcommand{\ODD}{\ensuremath{\texttt{Obj}_\texttt{DD}}}
\newcommand{\OUU}{\ensuremath{\texttt{Obj}_\texttt{UU}}}

\newcommand{\utiltp}{\ensuremath{\texttt{util}_{t,p}}}

\newcommand{\profitsjtp}{\ensuremath{\texttt{profits}_{j,t,p}}}
\newcommand{\profitstp}{\ensuremath{\texttt{profits}_{t,p}}}

\newcommand{\relocateddt}{\ensuremath{\texttt{relocated}_{d,t}}}
\newcommand{\overheadtd}{\ensuremath{\texttt{overhead}_{d,t}}}
\newcommand{\overheadtp}{\ensuremath{\texttt{overhead}_{t,d}}}
\newcommand{\overheadjt}{\ensuremath{\texttt{overhead}_{j,d}}}

\newcommand{\backupswitch}{\ensuremath{s_B}}

\newcommand{\Yrt}{\ensuremath{Y_{r,t}}}
\newcommand{\Yjrt}{\ensuremath{Y_{j \ArrowAllocate r,t}}}
\newcommand{\YjrtOpt}{\ensuremath{Y^*_{j \ArrowAllocate r,t}}}

\newcommand{\YjrtMinusOne}{\ensuremath{Y_{j \ArrowAllocate r,t_{-1}}}}
\newcommand{\YjrtCaseA}{\ensuremath{Y_{j \ArrowAllocate r_1,t_{-1}}}}
\newcommand{\YjrtCaseB}{\ensuremath{Y_{j \ArrowAllocate r_2,t}}}

\newcommand{\Xdt}{\ensuremath{X_{d,t}}} 
\newcommand{\XdtOpt}{\ensuremath{X_{d,t}^*}}
\newcommand{\XdA}{\ensuremath{X_{d,A}}}
\newcommand{\XdAOpt}{\ensuremath{X^*_{d,A}}}
\newcommand{\XdtZero}{\ensuremath{X^{\TZeroColor}_{d}}} 
\newcommand{\XdtZeroOpt}{\ensuremath{X^{\TZeroColor}_{d}}} 
\newcommand{\XdtOne}{\ensuremath{X^{\TOneColor}_{d}}} 
\newcommand{\XdtOneOpt}{\ensuremath{X^{\TOneColor}_{d}}}

\newcommand{\sumAj}{\ensuremath{\sum_{a \in A_j}}} 
\newcommand{\sumRjt}{\ensuremath{\sum_{r \in R_{j,t}}}}
\newcommand{\sumS}{\ensuremath{\sum_{s \in S}}} 
\newcommand{\sumSr}{\ensuremath{\sum_{r \in S}}}
\newcommand{\sumT}{\ensuremath{\sum_{t \in T}}} 
\newcommand{\sumZ}{\ensuremath{\sum_{s \in Z_{j,t}}}} 
\newcommand{\sumTj}{\ensuremath{\sum_{t \in T_j}}} 
\newcommand{\sumTi}{\ensuremath{\sum_{t \in T_i}}} 
\newcommand{\sumP}{\ensuremath{\sum_{p \in P}}} 
\newcommand{\sumD}{\ensuremath{\sum_{d \in \Dst}}} 
\newcommand{\sumDs}{\ensuremath{\sum_{d \in \Ds}}} 
\newcommand{\sumF}{\ensuremath{\sum_{f \in F}}} 
\newcommand{\sumFst}{\ensuremath{\sum_{f \in F_{s,t}}}} 
\newcommand{\sumFdt}{\ensuremath{\sum_{f \in \Fdt}}} 
\newcommand{\sumJ}{\ensuremath{\sum_{j \in J}}}
\newcommand{\sumJt}{\ensuremath{\sum_{j \in J_t}}}
\newcommand{\sumJT}{\ensuremath{\sum_{j \in J_T}}}
\newcommand{\sumAdT}{\ensuremath{\sum_{A \in \mathbb{A}_{d}}}} 
\newcommand{\sumAdTOpt}{\ensuremath{\sum_{A \in \mathbb{A}^*_{d}}}} 
\newcommand{\sumAjT}{\ensuremath{\sum_{A \in A_{j,T_j}}}} 
\newcommand{\sumAjTOpt}{\ensuremath{\sum_{A \in A^*_{j,T_j}}}} 

\newcommand{\allRjt}{\ensuremath{\displaystyle\mathop{\forall}_{r \in R_{j,t}}}}
\newcommand{\allS}{\ensuremath{\displaystyle\mathop{\forall}_{s \in S}}}
\newcommand{\allSr}{\ensuremath{\displaystyle\mathop{\forall}_{r \in S}}}
\newcommand{\allD}{\ensuremath{\displaystyle\mathop{\forall}_{d \in D}}}
\newcommand{\allDs}{\ensuremath{\displaystyle\mathop{\forall}_{d \in D_s}}}
\newcommand{\allDst}{\ensuremath{\displaystyle\mathop{\forall}_{d \in \Dst}}}
\newcommand{\allA}{\ensuremath{\displaystyle\mathop{\forall}_{a \in A_j}}}
\newcommand{\allAd}{\ensuremath{\displaystyle\mathop{\forall}_{a \in A_d}}}
\newcommand{\allAr}{\ensuremath{\displaystyle\mathop{\forall}_{r \in A_{j,t}}}}
\newcommand{\allZ}{\ensuremath{\displaystyle\mathop{\forall}_{s \in Z_{j,t}}}}
\newcommand{\allP}{\ensuremath{\displaystyle\mathop{\forall}_{p \in P}}}
\newcommand{\allI}{\ensuremath{\displaystyle\mathop{\forall}_{i \in I}}}
\newcommand{\allJ}{\ensuremath{\displaystyle\mathop{\forall}_{j \in J}}}
\newcommand{\allT}{\ensuremath{\displaystyle\mathop{\forall}_{t \in T}}}
\newcommand{\allTi}{\ensuremath{\displaystyle\mathop{\forall}_{t \in T_i}}}
\newcommand{\allTj}{\ensuremath{\displaystyle\mathop{\forall}_{t \in T_j}}}
\newcommand{\allJt}{\ensuremath{\displaystyle\mathop{\forall}_{j \in J_t}}}
\newcommand{\allJT}{\ensuremath{\displaystyle\mathop{\forall}_{j \in J_T}}}
\newcommand{\allAdT}{\ensuremath{\displaystyle\mathop{\forall}_{A \in \mathbb{A}_{d}}}}

\newcommand{\TLink}{\text{\tiny Link}}
\newcommand{\TTable}{\text{\tiny Table}}
\newcommand{\TStatic}{\text{\tiny Static}}
\newcommand{\TCtrl}{\text{\tiny Ctrl}}
\newcommand{\TDist}{\text{\tiny Dist}}
\newcommand{\TTableOverhead}{\text{\tiny Table}}
\newcommand{\TLinkOverhead}{\text{\tiny Link}}
\newcommand{\TCtrlOverhead}{\text{\tiny Ctrl}}
\newcommand{\TAlloc}{\texttt{\tiny RSA}}
\newcommand{\TSelect}{\texttt{\tiny DTS}}
\newcommand{\TZero}{\text{\rectangled{0}}}
\newcommand{\TOne}{\text{\rectangled{1}}}

\newcommand{\TZeroColor}{\text{\rectangledDef{0}}}
\newcommand{\TOneColor}{\text{\rectangledDef{1}}}

\newcommand{\TZeroZero}{\TZeroColor\TZero}
\newcommand{\TZeroOne}{\TZeroColor\TOne}
\newcommand{\TOneZero}{\TOneColor\TZero}
\newcommand{\TOneOne}{\TOneColor\TOne}

\newcommand{\caseZeroZeroC}{\text{\circled{1}}}
\newcommand{\caseZeroOneC}{\text{\circled{2}}}
\newcommand{\caseOneZeroC}{\text{\circled{3}}}
\newcommand{\caseOneOneC}{\text{\circled{4}}}

\newcommand{\caseZeroZero}{\small\text{$\TZeroZero$}}
\newcommand{\caseZeroOne}{\small\text{$\TZeroOne$}}
\newcommand{\caseOneZero}{\small\text{$\TOneZero$}}
\newcommand{\caseOneOne}{\small\text{$\TOneOne$}}

\newcommand{\caseZeroZeroW}{\small\text{$\TZero\TZero$}}
\newcommand{\caseZeroOneW}{\small\text{$\TZero\TOne$}}
\newcommand{\caseOneZeroW}{\small\text{$\TOne\TZero$}}
\newcommand{\caseOneOneW}{\small\text{$\TOne\TOne$}}

\newcommand{\ArrowAllocate}{,}
\newcommand{\ArrowLink}{,}

\newcommand{\cTableRaw}{\ensuremath{c^{\TTable}}} 
\newcommand{\cTable}{\ensuremath{c^{\TTable}_s}} 
\newcommand{\cTabler}{\ensuremath{c^{\TTable}_r}} 
\newcommand{\cTablea}{\ensuremath{c^{\TTable}_{a_t}}} 
\newcommand{\cTablerA}{\ensuremath{c^{\TTable}_{a_t}}} 
\newcommand{\uTable}{\ensuremath{u^{\TTable}_{s,t}}} 
\newcommand{\uTablest}{\uTable} 
\newcommand{\uTablestOne}{\ensuremath{u^{\TTable}_{s,t_1}}} 
\newcommand{\uTablert}{\ensuremath{u^{\TTable}_{r,t}}} 
\newcommand{\uTablertA}{\ensuremath{u^{\TTable}_{a_t,t}}} 
\newcommand{\uTablea}{\ensuremath{u^{\TTable}_{a_t,t}}}

\newcommand{\uLinkaj}{\ensuremath{u^{\textit{Link}}_{a_t,s_j,t}}} 
\newcommand{\uLinkja}{\ensuremath{u^{\textit{Link}}_{s_j,a_t,t}}} 


\newcommand{\wdt}{\ensuremath{w_{d,t}}}
\newcommand{\wdA}{\ensuremath{w_{d,A}}}
\newcommand{\wjrt}{\ensuremath{w_{j \ArrowAllocate r,t}}}
\newcommand{\wjrtA}{\ensuremath{w_{j \ArrowAllocate a_t,t}}}
\newcommand{\wjA}{\ensuremath{w_{j,A}}}

\newcommand{\wdZeroZero}{\ensuremath{w^{\tiny\TZeroZero}_{d}}} 
\newcommand{\wdZeroOne}{\ensuremath{w^{\tiny\TZeroOne}_{d}}} 
\newcommand{\wdOneZero}{\ensuremath{w^{\tiny\TOneZero}_{d}}} 
\newcommand{\wdOneOne}{\ensuremath{w^{\tiny\TOneOne}_{d}}}

\newcommand{\wStaticrt}{\ensuremath{w^{\TStatic}_{r,t}}}
\newcommand{\wTablejrt}{\ensuremath{w^{\TTable}_{j \ArrowAllocate r,t}}}
\newcommand{\wLinkjrt}{\ensuremath{w^{\TLink}_{j \ArrowAllocate r,t}}}
\newcommand{\wCtrljrt}{\ensuremath{w^{\TCtrl}_{j \ArrowAllocate r,t}}}
\newcommand{\wDistjrt}{\ensuremath{w^{\TDist}_{j \ArrowAllocate r,t}}}

\newcommand{\wStaticrtA}{\ensuremath{w^{\TStatic}_{a_t,t}}}
\newcommand{\wTablejrtA}{\ensuremath{w^{\TTable}_{j \ArrowAllocate a_t,t}}}
\newcommand{\wLinkjrtA}{\ensuremath{w^{\TLink}_{j \ArrowAllocate a_t,t}}}
\newcommand{\wCtrljrtA}{\ensuremath{w^{\TCtrl}_{j \ArrowAllocate a_t,t}}}
\newcommand{\wDistjrtA}{\ensuremath{w^{\TDist}_{j \ArrowAllocate a_t,t}}}

\newcommand{\wStaticA}{\ensuremath{w^{\TStatic}_{A}}}
\newcommand{\wTablejA}{\ensuremath{w^{\TTable}_{j,A}}}
\newcommand{\wLinkjA}{\ensuremath{w^{\TLink}_{j,A}}}
\newcommand{\wCtrljA}{\ensuremath{w^{\TCtrl}_{j,A}}}
\newcommand{\wDistjA}{\ensuremath{w^{\TDist}_{j,A}}}

\newcommand{\wLink}{\ensuremath{w^{\TLink}_{j,a}}} 
\newcommand{\wDist}{\ensuremath{w^{\mathit{Dist}}_{j,a}}} 
\newcommand{\wCtrl}{\ensuremath{w^{\mathit{Ctrl}}_{j,a}}} 

\newcommand{\WdtZeroZero}{\ensuremath{W^{\tiny\TZeroZero\TTableOverhead}_{d,t}}} 
\newcommand{\WdtZeroOne}{\ensuremath{W^{\tiny\TZeroOne\TTableOverhead}_{d,t}}}

\newcommand{\wTable}{\ensuremath{w^{\TTable}_{j,a}}} 
\newcommand{\wTabled}{\ensuremath{w^{\TTableOverhead}_{d}}} 
\newcommand{\wTabledt}{\ensuremath{w^{\TTableOverhead}_{d,t}}} 
\newcommand{\wTableda}{\ensuremath{w^{\TTableOverhead}_{d,~\!\!A}}} 
\newcommand{\wTabledta}{\ensuremath{w^{\TTableOverhead}_{d,t,~\!\!A}}}
\newcommand{\wTabledZero}{\ensuremath{w^{\tiny\TZero\TTableOverhead}_{d}}} 
\newcommand{\wTabledOne}{\ensuremath{w^{\tiny\TOne\TTableOverhead}_{d}}} 
\newcommand{\wTabledZeroZero}{\ensuremath{w^{\tiny\TZeroZero\TTableOverhead}_{d}}} 
\newcommand{\wTabledZeroOne}{\ensuremath{w^{\tiny\TZeroOne\TTableOverhead}_{d}}} 
\newcommand{\wTabledOneZero}{\ensuremath{w^{\tiny\TOneZero\TTableOverhead}_{d}}} 
\newcommand{\wTabledOneOne}{\ensuremath{w^{\tiny\TOneOne\TTableOverhead}_{d}}} 
\newcommand{\wTabledtaZeroZero}{\ensuremath{w^{\tiny\TZeroZero\TTableOverhead}_{d,t,~\!\!A}}} 
\newcommand{\wTabledtaZeroOne}{\ensuremath{w^{\tiny\TZeroOne\TTableOverhead}_{d,t,~\!\!A}}} 
\newcommand{\wTabledtaOneZero}{\ensuremath{w^{\tiny\TOneZero\TTableOverhead}_{d,t,~\!\!A}}} 
\newcommand{\wTabledtaOneOne}{\ensuremath{w^{\tiny\TOneOne\TTableOverhead}_{d,t,~\!\!A}}} 
\newcommand{\wTabledaZeroZero}{\ensuremath{w^{\tiny\TZeroZero\TTableOverhead}_{d,a}}} 
\newcommand{\wTabledaZeroOne}{\ensuremath{w^{\tiny\TZeroOne\TTableOverhead}_{d,a}}} 
\newcommand{\wTabledaOneZero}{\ensuremath{w^{\tiny\TOneZero\TTableOverhead}_{d,a}}} 
\newcommand{\wTabledaOneOne}{\ensuremath{w^{\tiny\TOneOne\TTableOverhead}_{d,a}}} 

\newcommand{\wLinkd}{\ensuremath{w^{\TLinkOverhead}_{d}}} 
\newcommand{\wLinkdt}{\ensuremath{w^{\TLinkOverhead}_{d,t}}} 
\newcommand{\wLinkda}{\ensuremath{w^{\TLinkOverhead}_{d,~\!\!A}}} 
\newcommand{\wLinkdZero}{\ensuremath{w^{\tiny\TZero\TLinkOverhead}_{d}}} 
\newcommand{\wLinkdOne}{\ensuremath{w^{\tiny\TOne\TLinkOverhead}_{d}}} 
\newcommand{\wLinkdZeroZero}{\ensuremath{w^{\tiny\TZeroZero\TLinkOverhead}_{d}}} 
\newcommand{\wLinkdZeroOne}{\ensuremath{w^{\tiny\TZeroOne\TLinkOverhead}_{d}}} 
\newcommand{\wLinkdOneZero}{\ensuremath{w^{\tiny\TOneZero\TLinkOverhead}_{d}}} 
\newcommand{\wLinkdOneOne}{\ensuremath{w^{\tiny\TOneOne\TLinkOverhead}_{d}}} 
\newcommand{\wLinkdtaZeroZero}{\ensuremath{w^{\tiny\TZeroZero\TLinkOverhead}_{d,t,~\!\!A}}} 
\newcommand{\wLinkdtaZeroOne}{\ensuremath{w^{\tiny\TZeroOne\TLinkOverhead}_{d,t,~\!\!A}}} 
\newcommand{\wLinkdtaOneZero}{\ensuremath{w^{\tiny\TOneZero\TLinkOverhead}_{d,t,~\!\!A}}} 
\newcommand{\wLinkdtaOneOne}{\ensuremath{w^{\tiny\TOneOne\TLinkOverhead}_{d,t,~\!\!A}}} 

\newcommand{\wCtrld}{\ensuremath{w^{\TCtrlOverhead}_{d}}} 
\newcommand{\wCtrldt}{\ensuremath{w^{\TCtrlOverhead}_{d,t}}} 
\newcommand{\wCtrlda}{\ensuremath{w^{\TCtrlOverhead}_{d,~\!\!A}}} 
\newcommand{\wCtrldZero}{\ensuremath{w^{\tiny\TZero\TCtrlOverhead}_{d}}} 
\newcommand{\wCtrldOne}{\ensuremath{w^{\tiny\TOne\TCtrlOverhead}_{d}}} 
\newcommand{\wCtrldZeroZero}{\ensuremath{w^{\tiny\TZeroZero\TCtrlOverhead}_{d}}} 
\newcommand{\wCtrldZeroOne}{\ensuremath{w^{\tiny\TZeroOne\TCtrlOverhead}_{d}}} 
\newcommand{\wCtrldOneZero}{\ensuremath{w^{\tiny\TOneZero\TCtrlOverhead}_{d}}} 
\newcommand{\wCtrldOneOne}{\ensuremath{w^{\tiny\TOneOne\TCtrlOverhead}_{d}}} 
\newcommand{\wCtrldtaZeroZero}{\ensuremath{w^{\tiny\TZeroZero\TCtrlOverhead}_{d,t,~\!\!A}}} 
\newcommand{\wCtrldtaZeroOne}{\ensuremath{w^{\tiny\TZeroOne\TCtrlOverhead}_{d,t,~\!\!A}}} 
\newcommand{\wCtrldtaOneZero}{\ensuremath{w^{\tiny\TOneZero\TCtrlOverhead}_{d,t,~\!\!A}}} 
\newcommand{\wCtrldtaOneOne}{\ensuremath{w^{\tiny\TOneOne\TCtrlOverhead}_{d,t,~\!\!A}}} 

\newcommand{\ujrt}{\ensuremath{u_{j \ArrowAllocate r,t}}}
\newcommand{\udta}{\ensuremath{u^{\TTable}_{d,t,~\!\!A}}}

\newcommand{\udtZero}{\ensuremath{u^{\TZero}_{d,t}}} 
\newcommand{\udtOne}{\ensuremath{u^{\TOne}_{d,t}}} 
\newcommand{\udtZeroZero}{\ensuremath{u^{\TZeroZero}_{d,t}}} 
\newcommand{\udtZeroOne}{\ensuremath{u^{\TZeroOne}_{d,t}}} 
\newcommand{\udtOneZero}{\ensuremath{u^{\TOneZero}_{d,t}}} 
\newcommand{\udtOneOne}{\ensuremath{u^{\TOneOne}_{d,t}}} 

\newcommand{\uLinksj}{\ensuremath{u^{\TLink}_{s \ArrowLink s_j,t}}} 
\newcommand{\uLinkjs}{\ensuremath{u^{\TLink}_{s_j \ArrowLink s,t}}} 
\newcommand{\uLinksr}{\ensuremath{u^{\TLink}_{s \ArrowLink r,t}}} 
\newcommand{\uLinkjr}{\ensuremath{u^{\TLink}_{s_j \ArrowLink r,t}}} 
\newcommand{\uLinkrj}{\ensuremath{u^{\TLink}_{r \ArrowLink s_j,t}}} 
\newcommand{\uLinkjrA}{\ensuremath{u^{\TLink}_{s_j \ArrowLink a_t,t}}} 
\newcommand{\uLinkrjA}{\ensuremath{u^{\TLink}_{a_t \ArrowLink s_j,t}}} 

\newcommand{\dLink}{\ensuremath{u^{\textit{Link}}_{j,t}}} 
\newcommand{\dLinkjt}{\ensuremath{u^{\TLink}_{j,t}}}
\newcommand{\dTablejt}{\ensuremath{u^{\TTable}_{j,t}}} 
\newcommand{\dTable}{\ensuremath{u^{\TTable}_{j,t}}} 

\newcommand{\omegaDTS}{\ensuremath{\omega_\TSelect}} 
\newcommand{\omegaRSA}{\ensuremath{\omega_\TAlloc}} 
\newcommand{\omegaTableDTS}{\ensuremath{\omega^{\TTable}_\TSelect}} 
\newcommand{\omegaLinkDTS}{\ensuremath{\omega^{\TLink}_\TSelect}} 
\newcommand{\omegaCtrlDTS}{\ensuremath{\omega^{\TCtrl}_\TSelect}} 

\newcommand{\omegaTableRSA}{\ensuremath{\omega^{\TTable}_\TAlloc}} 
\newcommand{\omegaLinkRSA}{\ensuremath{\omega^{\TLink}_\TAlloc}} 
\newcommand{\omegaCtrlRSA}{\ensuremath{\omega^{\TCtrl}_\TAlloc}} 
\newcommand{\omegaDistRSA}{\ensuremath{\omega^{\TDist}_\TAlloc}}
\newcommand{\omegaStaticRSA}{\ensuremath{\omega^{\TStatic}_\TAlloc}}

\newcommand{\cLink}{\ensuremath{c^{\TLink}}} 
\newcommand{\cLinksj}{\ensuremath{c^{\TLink}_{s \ArrowLink s_j}}} 
\newcommand{\cLinkjs}{\ensuremath{c^{\TLink}_{s_j \ArrowLink s}}} 
\newcommand{\cLinkjr}{\ensuremath{c^{\TLink}_{s_j \ArrowLink r}}} 
\newcommand{\cLinkrj}{\ensuremath{c^{\TLink}_{r \ArrowLink s_j}}} 
\newcommand{\cLinkjrA}{\ensuremath{c^{\TLink}_{s_j \ArrowLink a_t}}} 
\newcommand{\cLinkrjA}{\ensuremath{c^{\TLink}_{a_t \ArrowLink s_j}}} 
\newcommand{\cLinkaj}{\ensuremath{c^{\TLink}_{a_t,s_j}}} 
\newcommand{\cLinkja}{\ensuremath{c^{\TLink}_{s_j,a_t}}} 
\newcommand{\cLinksr}{\ensuremath{c^{\TLink}_{s \ArrowLink r}}}

\newcommand{\HdXColor}{\ensuremath{\text{\rectangledDef{$H^X_d$}}}}

\newcommand{\HdX}{\ensuremath{H^X_d}}
\newcommand{\HdF}{\ensuremath{H^F_d}}

\newcommand{\AdT}{\ensuremath{\mathbb{A}_{d}}} 
\newcommand{\AdTOpt}{\ensuremath{\mathbb{A}^*_{d}}} 
\newcommand{\AjT}{\ensuremath{A_{j,T_j}}} 
\newcommand{\AjTOpt}{\ensuremath{A^*_{j,T_j}}}

\newcommand{\Tij}{\ensuremath{\hat{T}_{i,j}}} 
\newcommand{\Fst}{\ensuremath{F_{s,t}}} 
\newcommand{\Fjt}{\ensuremath{F_{j,t}}}

\newcommand{\Mst}{\ensuremath{M_{s,t}}} 
\newcommand{\Ds}{\ensuremath{D_{s}}} 
\newcommand{\Dst}{\ensuremath{D_{s,t}}} 
\newcommand{\DsT}{\ensuremath{D_{s,T}}} 
\newcommand{\ysr}{\ensuremath{y_{s \ArrowLink r}}} 
\newcommand{\yrs}{\ensuremath{y_{r \ArrowLink s}}} 
\newcommand{\yhs}{\ensuremath{y_{h \ArrowLink s}}} 
\newcommand{\ysh}{\ensuremath{y_{s \ArrowLink h}}} 

\newcommand{\FstCS}{\ensuremath{\Fst^\CS}}
\newcommand{\Fd}{\ensuremath{F_d}}
\newcommand{\Rjt}{\ensuremath{R_{j,t}}}
\newcommand{\udt}{\ensuremath{u^{\TTable}_{d,t}}}

\newcommand{\Fdt}{\ensuremath{F_{d,t}}}
\newcommand{\md}{\ensuremath{\overrightarrow{m_d}}}
\newcommand{\mj}{\ensuremath{\overrightarrow{m_j}}}

\newcommand{\prio}{\ensuremath{\texttt{prio}}}
\newcommand{\prioagg}{\ensuremath{\texttt{prio}_\texttt{agg}}}
\newcommand{\prioaggHigh}{\ensuremath{\prioagg^\texttt{highest}}}
\newcommand{\prioaggLow}{\ensuremath{\prioagg^\texttt{lowest}}}

\newcommand{\priod}{\ensuremath{\texttt{prio}_d}}
\newcommand{\fagg}{\ensuremath{f_\texttt{agg}}}
\newcommand{\magg}{\ensuremath{\overrightarrow{m_\texttt{agg}}}}
\newcommand{\aagg}{\ensuremath{\overrightarrow{a_\texttt{agg}}}}

\newcommand{\deltaft}{\ensuremath{\delta_{f,t}}} 

\newcommand{\lambdaA}{\ensuremath{\lambda^a_{f,t,d}}} 
\newcommand{\lambdaS}{\ensuremath{\lambda^i_{f,t,d}}} 
\newcommand{\lambdaI}{\ensuremath{\lambda^i_{f,t,d}}} 

\newcommand{\lambdaa}{\ensuremath{\lambda^a_{f,t}}} 
\newcommand{\lambdai}{\ensuremath{\lambda^i_{f,t}}} 

\newcommand{\phiK}{\ensuremath{\phi_{f,p}}} 
\newcommand{\phir}{\ensuremath{\phi_{f,r}}} 
\newcommand{\tinT}{\ensuremath{t \in T}} 
\newcommand{\pinP}{\ensuremath{p \in P}}

\newcommand{\evalGapvRLvUUvUU}{0.22}

\newcommand{\sE}{\ensuremath{\mathcal{E}}} 
\newcommand{\sT}{\ensuremath{\mathcal{T}}} 


\newcommand{\vsrc}{\ensuremath{v_\text{src}}}
\newcommand{\vdst}{\ensuremath{v_\text{dst}}}
\newcommand{\vctrl}{\ensuremath{v_\text{ctrl}}}

%% file: sections/introduction.tex

\section{Introduction}
\label{sec:intro}

%
Limited flow table capacity is a well known scalability and performance limitation for Software-defined Networking (SDN) that was intensively studied in the past \cite{CMTY+11-devoflow, YeTG13, JaMD14, KARW14-cacheflow}. But despite the fact that SDN is around for more than a decade, the situation has not yet changed significantly. Current hardware switches are still limited to a couple of thousand flow table entries and  capacity-related constraints are still present \cite{sahid19, flowmap19, Li2019}.
Flow delegation \cite{own-pbce, own-fda} is a recent concept to address flow table capacity bottlenecks by automatically relocating flow rules from a bottlenecked switch to neighboring switches with spare capacity.
Flow delegation can be used on-demand in a transparent fashion, without changes to network applications or other parts of the infrastructure, and is therefore well suited for scenarios where bottlenecks only occur in parts of the network or within certain time frames.

However, the benefits that can be achieved with flow delegation heavily depend on the used delegation algorithm. \cite{own-pbce} uses a simple greedy strategy based on two thresholds. This approach is inflexible in the sense that it cannot be optimized towards specific goals such as low link or low control overhead. 
\cite{own-fda} proposes a more flexible delegation algorithm that uses optimization based on a global target utilization value. However, it is based on the simplified assumption that all flow rules can be delegated individually. 
Potential conflicts between individual flow rules are not considered here. 

In this paper, we propose a delegation algorithm that makes use of a novel abstraction called delegation templates. This abstraction groups flow rules into sets that can be delegated without conflicts. Furthermore, we designed a multi-period problem formulation that can incorporate estimates of future network situations. It can also be simply extended with new objectives which is difficult in \cite{own-fda}.
We apply the new algorithm to almost 5000 different bottleneck scenarios with various
different characteristics and show that it can mitigate bottlenecks which
exceed the maximum flow table capacity by up to 38\% in the median case. This means we can
handle situations with 1380 concurrently installed flow rules while all switches in the
infrastructure have a capacity of only 1000 rules.
We can also use it to reduce operational and capital expenditure by buying switches with smaller flow tables: 28\% reduction in TCAM size in the median case which can lead to significant savings.
In addtion, our algorithm can be executed in less than 150ms per switch using one CPU core which means a single commodity server can handle a network with hundreds of switches.
The main contributions of this paper are:

\begin{itemize}
  \item A new delegation template abstraction to be used for flow delegation and an easy strategy to calculate delegation templates.
  \item A novel two-step algorithm for flow delegation based on multi-period multi-objective combinatorial optimization that takes predicted future network situations into account and can proactively mitigate anticipated bottlenecks.
  \item Evaluation of feasibility, performance, overhead, and runtime based on a wide range of different scenarios.
\end{itemize}

This paper is based on \cite{own-diss} but presents the results in a much more condensed format. Please refer to \cite{own-diss} for more detailed explanations and additional evaluation results. 
The remainder of this document is structured as follows. Sec.~\ref{sec:intro:approach:example} explains how flow delegation works. Sec.~\ref{sec:templates} introduces the delegation template abstraction.  Sec.~\ref{sec:algo} presents the new algorithm for flow delegation. Sec.~\ref{sec:eval} evaluates the proposed algorithm based on 4996 different bottleneck scenarios. 
Section~\ref{sec:rw} discusses related work. 
Finally, we give our concluding remarks in Section~\ref{sec:conclusion}.

%% file: sections/example.tex
\section{Background}
\label{sec:intro:approach:example}

\begin{figure}[tb!]
  \centering
  \includegraphics*[width=1\columnwidth]{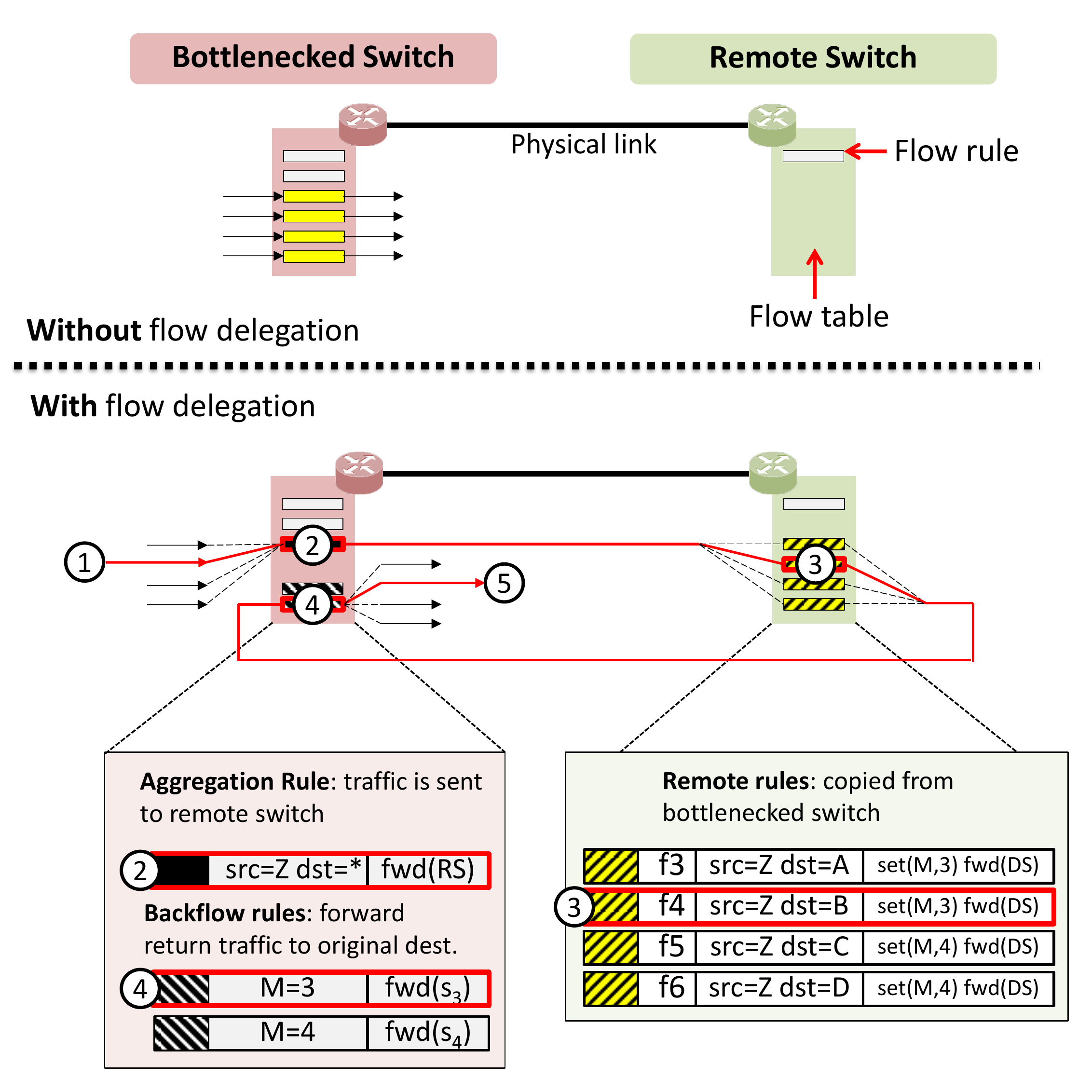}
  \caption{How flow delegation works}
  \label{fig:part1:process}
\end{figure}

The flow delegation concept is described in detail in \cite{own-pbce} and \cite{own-diss}. This section will briefly elaborate the underlying core mechanism.
Consider the example in Fig. \ref{fig:part1:process} where we assume that the flow table of the red switch on the left has a capacity of only six rules.
Without flow delegation, the red switch suffers from a flow table capacity bottleneck and can not handle any more rules (shown in the top of the figure).

However, the bottlenecked switch has a direct  physical link to another SDN switch that does not use all of its flow table capacity -- the so-called remote switch in green on the right hand of the figure.
Flow delegation will now relocate some of the flow rules from the left to the right to avoid the bottleneck.
The bottom part of the figure shows how this is done.
To make sure that flow delegation does not interfere with the logic of the network application, it is required that all flow rules ``remain as they are'', i.e., all packets are processed exactly as intended by the network application. For this to work, a detour procedure is introduced that makes use of three classes of flow rules \cite{own-pbce}: aggregation rules, remote rules, and backflow rules. 

\textbf{Aggregation rules} are installed in the bottlenecked switch in order to forward traffic to the remote switch. 
 They use wildcard matches that ``cover`` a subset  of the flow rules in the flow table.
 Every packet matched by a rule in the cover set is also matched by the aggregation rule.
 Consider $f_3$ to $f_6$ in the example. 
 Now consider an aggregation rule that matches all packets sent from source Z regardless of the destination. This can be expressed with a wildcard for the destination part of the match, i.e., $\texttt{src} = Z, \texttt{dst} = *$. It is easy to see that every packet that has to be processed by one of the yellow rules is also matched by this new wildcard rule. 

\textbf{Remote rules} are installed in the remote switch. They represent the ''delegated`` rules  that were relocated to the remote switch. They are basically a copy of the original rule from the bottlenecked switch, with one small exception: The action part of the rule is modified so that i) a marker is attached to the packet that encodes the forwarding decision of the network application and ii) packets are sent back to the bottlenecked switch after processing.

\textbf{Backflow rules} are installed in the bottlenecked switch. They are used to forward the marked packets from a remote rule towards their correct destination. Because the maximum number of potential forwarding destinations is limited by the number of interfaces of the bottlenecked switch, the number of backflow rules is also strictly limited to the same value.

Using these three classes of rules, the detour procedure is realized as follows: It first takes a set of rules from the bottlenecked switch (e.g., the four yellow rules $f_3$ to $f_6$) that can be covered by an aggregation rule. The rules from this cover set are transformed into remote rules and installed in the remote switch. Next, the aggregation and backflow rules are  installed in the bottlenecked switch. Finally, the rules from the cover set are removed from the bottlenecked switch which will decrease its flow table utilization. In this case, four rules are removed from the bottlenecked switch. Because one aggregation rule and two backflow rules are required, the total utilization is only reduced by 1 in this trivial example. This is different in more realistic scenarios where the cover set can contain hundreds of rules while the number of aggregation and backflow rules stays small.

So what happens when a packet arrives at the bottlenecked switch when some of the rules are delegated? This is illustrated by steps \circled{1} to \circled{5} in Fig. \ref{fig:part1:process}.
Assume the packet must be processed by flow rule $f_4$ that was relocated together with the other yellow flow rules to the remote switch. The four yellow rules were replaced with an aggregation rule (match: $\texttt{src}=Z, \texttt{dst}=*$). So the packet is matched by the aggregation rule and sent towards the remote switch in step \circled{2}. After the packet arrived at the remote switch it is matched by the remote rule for $f_4$ (match: $\texttt{src}=Z, \texttt{dst}=B$) in step \circled{3}. The forwarding decision of the original rule is attached as a marker with $\texttt{set}(M, 3)$ and the packet is sent back to the bottlenecked switch. In step \circled{4}, one of the backflow rules (match: $M=3$) is matched and the packet is finally forwarded to the intended destination in step \circled{5}.

%% file: sections/functionality.tex

\section{Delegation Templates}
\label{sec:templates}

Delegation templates are a new abstraction designed for flow delegation -- and potentially also for similar tasks.
In essence, a delegation template $d \in \Dst$ for switch $s$ represents a cover set $\Fst^\CS$ of flow rules which can be relocated to a remote switch without rule conflicts. 
More precisely, it is guaranteed that no rule conflicts occur if all flow rules in $\Fst^\CS$ are kept together, i.e., all of them are either installed in the remote switch or in the bottlenecked switch.
Time index  $t$ is required because flow rules within switch $s$ change over time.
We use this construct to model delegation decisions as a selection problem. If a delegation template is not selected, nothing happens. If a delegation template is selected, an aggregation rule $\magg$ is created that covers all rules in $\Fst^\CS$ and redirects the associated traffic to the remote switch. This match is  installed into the bottlenecked switch and the flow table utilization will be decreased by the cardinality of the cover set. 
Sec. \ref{sec:templates-example} further explains the concept using a small example. 
Sec. \ref{sec:templates-calculation} then elaborates how $\Dst$ can be created in practice.

\subsection{Example}
\label{sec:templates-example}

\begin{figure}[htb!]
  \centering
  \includegraphics*[width=\columnwidth]{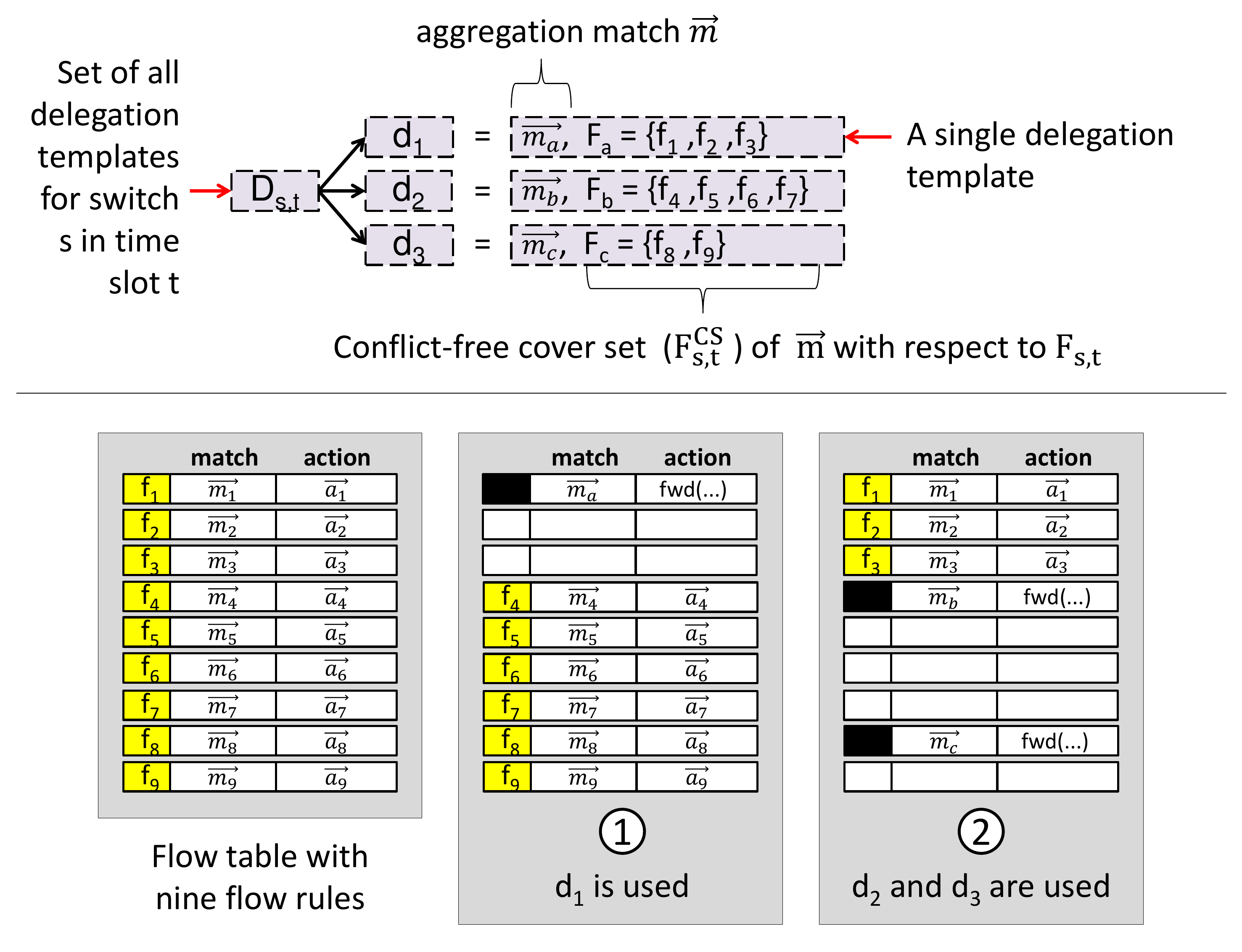}
  \caption{Delegation template example}
  \label{fig:delegation-template-example}
\end{figure}

Fig \ref{fig:delegation-template-example} shows a flow table with capacity $\cTable = 9$ and nine installed flow rules $\Fst := \{f_1, \ldots, f_9\}$. The figure further shows three delegation templates given as set $\Dst$. 
Template $d_1 \in \Dst$ consists of match $\overrightarrow{m_\texttt{a}}$ and cover set $F_a = \{f_1, f_2, f_3\}$. $F_a$ is calculated by Alg. \ref{algo:cover-set} based on $\overrightarrow{m_\texttt{a}}$.
The flow table in the middle labeled as \circled{1} now shows what happens if delegation template $d_1$ would be selected for flow delegation. 
First, the flow rules  $f_1$, $f_2$, and $f_3$ are relocated to the remote switch (not shown). Afterwards, a new aggregation rule with match $\overrightarrow{m_\texttt{a}}$ is installed that forwards the traffic for $f_1$, $f_2$, and $f_3$ to the remote switch (shown in black). At the same time, the three rules  $f_1$, $f_2$, and $f_3$ are removed and the utilization of the flow table is reduced by $|F_a| - 1 = 2$.
The other delegation templates $d_2$ and $d_3$ represent two different options. The flow table in the right depicts the situation if $d_2$ and $d_3$ are selected instead of $d_1$. In this case, there are two aggregation rules: one for $d_2$ with match $\overrightarrow{m_\texttt{b}}$ and another one for $d_3$ with $\overrightarrow{m_\texttt{c}}$.
With this selection, utilization of the flow table is reduced by $|F_b| + |F_c| - 2 = 4$ (the $-2$ represents the two required aggregation rules).

\vvv

\begin{algorithm}[htb!]
    \DontPrintSemicolon
    \SetNoFillComment
    \SetKwRepeat{Assert}{assert}{end}
    \SetKwFunction{FMain}{cover\_set}
    \SetKwProg{Fn}{Function}{:}{end}
    
    \KwData{aggregation match $\magg$, aggregation priority $\prioagg$, set of flow rules $\Fst$ }
    \KwResult{conflict-free cover set $\Fst^\CS$ for $\magg$ with respect to $\Fst$}
      
    \Fn{\FMain{$\magg$,~ $\Fst$,~ $\prioagg$}}{
        $\Fst^\CS \setTo \{\}$\;
        $F^\texttt{sorted} \setTo \text{sort}\big(\Fst \setminus \big\{f_\texttt{default} \big\} \big)$ \label{algo:cover-set-sort}\;
        $Z_\texttt{agg} \setTo $ all packets matched by $\overrightarrow{m_\texttt{agg}}$\;
        \For{$f_i = \big<\overrightarrow{m_i}, \overrightarrow{a_i}, \prio_i \big> \in F^\texttt{sorted}$}{
            \If{$\prio_\texttt{agg} \geq \prio_i$}{
                $Z_i \setTo $ all packets matched by $\overrightarrow{m_i}$\;
                
                $Z_\texttt{intersect} \setTo Z_\texttt{agg} \cap Z_i$\;
                
                \If{$Z_\texttt{intersect} \neq \emptyset$}{ \label{algo:cover-set-intersect}
                    $\Fst^\CS \setTo \Fst^\CS \cup \{f_i\}$\;
                    $Z_\texttt{agg} \setTo \big(Z_\texttt{agg} \setminus Z_\texttt{intersect} \big) \cup \big(Z_i \setminus Z_\texttt{intersect} \big)$ \label{algo:cover-set-update}\;
                }               
            }   
        }
        \Return $\Fst^\CS$
    }
 	\caption{Conflict-free cover set (based on \cite{rule-caching-algos-14})}
	\label{algo:cover-set}
\end{algorithm}

The example above shows only two valid possible selections (out of seven in this scenarios). 
The basic behavior, however, remains the same. 
The major benefit of the delegation template abstraction is separation of concerns. 
It allows it to develop rule aggregation schemes completely independent from the delegation process. 
This, in turn, makes it easy to benefit from existing algorithms for rule aggregation.


\subsection{Calculating Delegation Templates}
\label{sec:templates-calculation}

Given the abstraction from the previous section, we now need a way to actually calculate sets of delegation templates that can be used for mitigation. 
An easy way to do this is exploiting already existing structures, especially if these structures are designed in a conflict-free way. Assume systems such as FlowVisor \cite{flowvisor} or Covisor \cite{covisor} that create isolated slices for different controllers or network applications. Because of the inherent isolation requirement, delegation templates can be easily constructed based on the individual slices because rule conflicts across slices would violate the isolation requirement. 
However, this approach does not work with arbitrary network applications and requires changes in the controller. 
We therefore use a new indirect way to calculate delegation templates using additional context information.
$\Dst$ is not derived on-demand from the current set of installed flow rules $\Fst$. Instead, $\Dst$ is build ''indirectly`` as an independent set.
This has the (major) benefit that delegation templates can be defined in a static way without complex calculations.

Problem is that the main property of the delegation template abstraction must still be fulfilled , i.e., we need  a conflict-free cover set $F_{i,t}$  for an aggregation match $\overrightarrow{m_i}$ for each $d_i = \big<\overrightarrow{m_i}, F_{i,t} \big> \in \Dst$.
It can be shown that we have to guarantee the following four prerequisites in order to achieve this:

\begin{itemize}
 \item[(R1)] A set of $n$ delegation templates $\Dst := \{d_1, \ldots, d_n\}$ can be defined independently of the installed flow rules $\Fst$
 \item[(R2)] Each packet $z \in \mathcal{Z}$ can be linked to exactly one delegation template and the delegation template for all packets $z \in \mathcal{Z}$ can be inferred from packet $z$ using the same set of packet header fields
 \item[(R3)] Each rule in $f \in \Fst$ can be linked to exactly one delegation template and the delegation template for all rules $f \in \Fst$ can be inferred from rule $f$
 \item[(R4)] A packet $z \in \mathcal{Z}$ linked to delegation template $d_i$ is always processed by a rule f that is also linked to delegation template $d_i$
\end{itemize}

For the reminder of this paper, we use the so-called flow-to-ingress-port mapping presented in \cite{own-pbce} to define sets of delegation templates.
The key idea here is as follows: each physical port $p_i \in P_s$ of switch $s$ defines a delegation template $d_i \in \Dst$ and flow rules processing traffic that arrives at port $p_i$ are associated with delegation template $d_i$. 
In this scheme, R1 is fulfilled because the delegation templates can be derived from the number of physical ingress ports of a switch. R2 is fulfilled because a single packet $z$ can only arrive at one ingress port.
The two challenging prerequisites are R3 (each rule can be linked to exactly one delegation template) and R4 (a packet linked to delegation template $d_i$ is actually processed by a rule that is linked to delegation template $d_i$). 
Main problem: While the ingress port can be used in a match, the controller and the network applications are not forced to do so, i.e., the match for \texttt{in\_port} can also be set to a wildcard.
Such rules will violate prerequisite R3 because it is not possible to determine the corresponding delegation template. It also violates R4 because a packet with ingress port $p_i$ might be processed by a higher priority rule with a wildcard match for \texttt{in\_port}.

\vspace{0.8em}
This issue cannot be easily solved but it is possible to mitigate the effects to a large extend using information from context -- in this case, from the $\texttt{packet\_in}(z, p_i)$ control messages. We distinguishes between three cases:

\begin{enumerate}
 \item Rules that do match on a single \texttt{in\_port} $p_i$ are linked to delegation template $d_i$
 \item For reactively installed flow rules that do not match on \texttt{in\_port} the delegation template is inferred from the ingress port specified in the $\texttt{packet\_in}(z, p_i)$ control message that was sent to the controller.
 \item For proactive rules and reactive rules where the match on \texttt{in\_port} is explicitly set to a wildcard, an additional delegation template $d_0$ is introduced that represents rules that cannot be relocated. 
\end{enumerate}

Note that we cannot fully guarantee R4 this way because i) there can be rule conflicts between $d_0$ and one of the other delegation templates $d_1$ to $d_n$ and ii) inferring the delegation template for reactively installed flow rules is not necessarily 100\% accurate. So it is required to make use of rule conflict detection. However, this will not change the main benefit of the approach -- which is the static number of delegation templates that are independent from $\Fst$.

\section{Flow Delegation Algorithm}
\label{sec:algo}

\begin{figure}[htb!]
  \centering
  \includegraphics*[width=0.9\columnwidth]{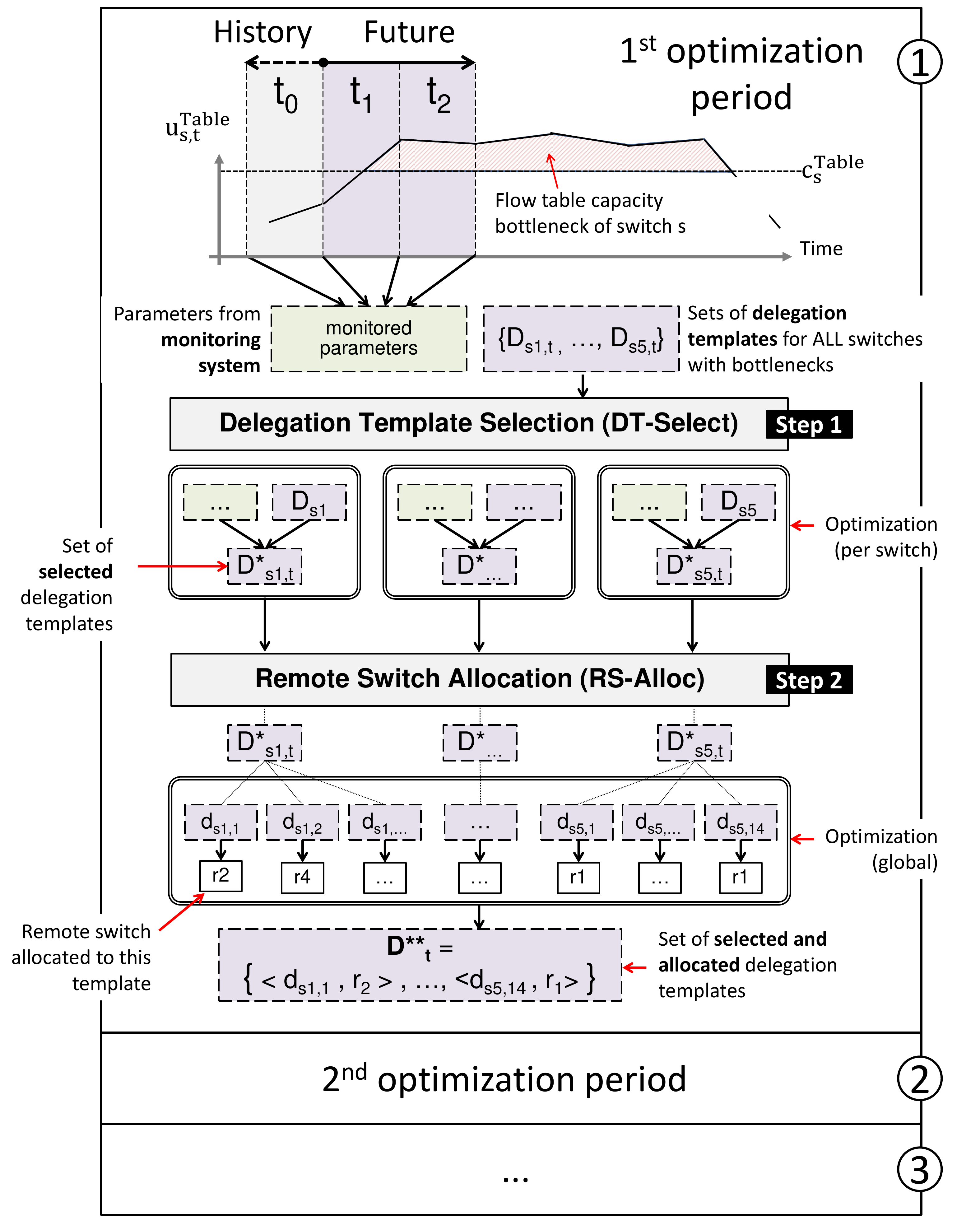}
  \caption{Periodic Two-Step Optimization Approach}
  \label{fig:da:decomposition}
\end{figure}

Based on the delegation template abstraction, flow delegation can be modeled as a periodic two-step optimization problem as shown in Fig. \ref{fig:da:decomposition}.
We assume that the flow delegation algorithm is executed periodically once per second as long as there are bottlenecks. It can be run on the same server as the SDN controller.
The individual executions are called optimization periods. In each optimization period, a limited future horizon of $|T|$ time slots is considered together with the status from the last optimization period referred to as history $H$. 

The three boxes labeled as \circled{1}, \circled{2} and \circled{3} in Fig. \ref{fig:da:decomposition} are three subsequent optimization periods. The two steps of the algorithm are executed in each period. The curve in the top represents raw data in form of the flow table utilization over time of a single switch $s$ estimated at the beginning of the first optimization period. 
This data comes from an existing network monitoring solution and is required to calculate the coefficients for the two optimization problems used in the following. 
The second part of input variables are the sets of delegation templates $\Dst$  for each bottlenecked switch calculated according to Sec. \ref{sec:templates-calculation} for all time slots that are included in the future horizon.


The first step now is selecting a subset of the delegation templates to mitigate the bottleneck(s) in the current time slot $t$. This is the Delegation Template Selection problem (DT-Select) which is solved independently for each bottlenecked switch.
The second step -- the allocation part -- is referred to as Remote Switch Allocation (RS-Alloc) and takes the set of selected delegation templates $\Dst^*$ for all bottlenecked switches and assigns each of it a remote switch based on flow table utilization and  utilization of the links in the infrastructure. The second step is executed only once per optimization period and returns a set of selected and assigned delegation templates $D_t^{**}$. This set is given to the SDN controller to implement the calculated delegation which will resolve the bottleneck.
Both, DT-Select and RS-Alloc are explained in more detail below.

\subsection{Step 1: DT-Select}
\label{sec:da:dts:modeling}

Goal of DT-Select is it to selects a subset $\Dst^* \subseteq \Dst$ out of $n$ delegation templates for switch $s$ so that i) the utilization of the flow table of switch $s$ does not exceed its capacity and ii) overhead caused by flow delegation is minimized.
This can be written as the following multi-period 0-1 knapsack problem:

\begin{IEEEeqnarray}{rCllcc}
\IEEEyesnumber
\min_{\Xdt}  &\quad& \sumD \sumT \Xdt * w_{d,t}  &  \label{eq:dts-mp}\\
\text{s.t.} &     & \uTable - \sumD \Xdt * \udt \le \cTable & \quad\quad & \allT  \label{eq:dts-mp:constraint}\\
            &     & \Xdt \in \{0,1\} &  &  \allDst & ~~\allT \label{eq:dts-mp:binary}
\end{IEEEeqnarray}

The knapsack is the flow table of bottlenecked switch $s$ and the items are the delegation templates $d \in \Dst$. 
The knapsack capacity is modeled as a negative value that represents the amount of flow rules that have to be relocated in order to mitigate the bottleneck.
The weight of each template is given as a utilization coefficient $\udt$ that represents the amount of rules that will be relocated if template $d$ is added to the knapsack.
And the cost of each template is given as a cost coefficient $w_{d,t}$ that represents the overhead associated with the template, e.g., how much traffic is processed by the relocated flow rules.
Goal is then to add delegation templates to the knapsack so that the cost is minimal and the capacity is non-negative. 
The iteration over multiple time slots ($t \in T$) is required because we want to take prediction of future network situations into account.

Given that the items in the knapsack  could be added and removed independently from each other for different time slots, this would be fairly simple to solve. Unfortunately, both coefficients ($\udt$ and $w_{d,t}$) are time dependent, i.e., to calculate the coefficients for time slot $t+1$, it is not enough to look at the situation at $t+1$. If $d$ is also selected in time slot $t$, we have to consider all flow rules that were installed at this earlier time slot and are still active at $t+1$. This continues for all $t \in T$ which forms a polynomial of degree $|T|$ which requires non-linear programming to be solved.
We therefore first transform this into a multi-dimensional multiple-choice knapsack problem as follows:

\begin{IEEEeqnarray}{rClllr}
\IEEEyesnumber
\min_{\XdA}  &\quad& \sumDs \sumAdT \XdA * \wdA  &   \label{eq:dssel:ilp:obj}\\
\text{s.t.} &	  & \uTable -  \sumDs \sumAdT  \XdA * \udta \le \cTable \quad\quad & \allT \label{eq:dssel:ilp:capacity-constraint} \\
 & & \sumAdT \XdA = 1  & \allDs  \label{eq:dssel:ilp:selectOne}\\
            &     & \XdA \in \{0,1\} & \allDs~\allAdT~\label{eq:dssel:ilp:binary}
\end{IEEEeqnarray}

Here, we have $n$ different mutually disjoint assignment sets $\AdT = \{A_1, \ldots, A_n\}$. Each delegation template defines one assignment set and each of these sets contains $2^m$ assignments. For an assignment $A \in \AdT$ and each time slot $t \in T$, there are coefficient $\udta$ and $\wdA$.
Goal is to choose exactly one item out of each assignment set so that the costs are minimized while the utilization is below the capacity in all time slots. Eq. (\ref{eq:dssel:ilp:obj}) ensures that the objective function is minimized. Eq. (\ref{eq:dssel:ilp:capacity-constraint}) models the capacity constraint. This is done for each time slot separately, i.e., there are $m=|T|$ capacity constraints.  The multiple-choice constraints in Eq. (\ref{eq:dssel:ilp:selectOne}) makes sure that exactly one out of $2^m$ assignments is chosen. 

Note that this is only an intermediate formulation that is even harder to solve than before because the assignment-based formulation grows exponentially with $m$. However, it allows us to define a simple heuristic where we  restrict the number of assignments considered in the decision problem. 
Our proposed heuristic for DT-Select restricts the delegation template selection to the first time slot $t_1$ while the remaining time slots $t_2, \ldots, t_m$ copy the decision from $t_1$. 
In other words: if delegation template $d$ is selected in $t_1$, this template is selected for all $t \in T$ and vice versa if the template is not selected.
The utilization and cost coefficients, however, are still calculated for all time slots, i.e., the utilization constraints must be fulfilled in all time slots and the cost is also minimized over all time slots. 
With this simplification, we can re-write the problem as a two-dimensional multiple-choice knapsack problem as follows:

\begin{IEEEeqnarray}{rClllr}
\IEEEyesnumber
\min_{\XdtZero, \XdtOne}  &\quad& \sumDs \bigg( (1-\HdX) * ~ \XdtZero     &  \\
\IEEEnonumber
 & & * ~ \Big( \wTabledZeroOne + \wLinkdZeroOne + \wCtrldZeroOne \Big) &  \\
\IEEEnonumber
 & & + ~ \HdX * ~  \XdtOne * \big( \wLinkdOneOne + \wCtrldOneOne \big)   &  \\
 \IEEEnonumber
 & & + ~ \HdX * ~ (1-\XdtOne) * \wCtrldOneZero  ~\bigg) &   \label{eq:ilp:idssrl}\\
\IEEEnonumber
 & &   &  \label{eq:ilp:idssrl:constraint}\\
\IEEEyesnumber
\text{s.t.} & & \uTable - \sumDs \bigg(~ (1-\HdX) * \XdtZero * \udtZeroOne  \\
\IEEEnonumber
 & & \quad\quad + ~ \HdX *  \XdtOne * \udtOneOne~ \bigg) \le  \cTable  & \allT \label{eq:ilp:idssrl:constraint2}  \\
            &     & \XdtZero, \XdtOne \in \{0,1\} & \allDs
\end{IEEEeqnarray}

Note that the decision variables do not have a time slot index because they are by design only associated with the first time slot in $T$. 
There are only four cases for the selection of delegation template $d$ based on the history of the last optimization period (given as $\big< H_d^X, H_d^F \big>$).
If $\HdX$ is set to $\TZeroColor$ (template $d$ was not selected in the previous optimization period), the decision variable for time slot $t_1$ can either be $\TZero$ or $\TOne$. And the same two options exist for $\HdX = \TOneColor$. Note that the value of $\HdX$ is displayed in grey because this value represents the history of the last optimization period and can not be changed any more.

The coefficients for this problem are calculated as follows. 
We use several binary helpers below which need to be derived from a network monitoring solution. $\lambdaa \in \{0,1\}$ is set to $1$ for all time slots where flow rule $f$ is active in the flow table and $\lambdai \in \{0,1\}$ is set to $1$ only for the one time slot where the flow rule is installed.
The utilization coefficients represent the amount of flow rules that will be relocated to the remote switch (``weight'' in the knapsack terminology). 
In case the template is not selected, this coefficient is set to zero. If the template is selected, there are two cases. If the template was not selected in the previous time slot, we can simply count the newly installed rules as:
\begin{equation}
        \udtZeroOne := \sum_{q=t_1}^{t} \sum_{F_{d,q}} \lambdai
\end{equation}

If the template was already selected, we have to take still active flow rules from recent history into consideration. All flow rules that were relocated to the remote switch prior to the current optimization period need to be included which is done by iterating over $H_{d}^F$. This is based on the assumption that aggregation rules are installed with low priority and pre-existing rules are not relocated.
\begin{equation}
        \udtOneOne := \sum_{q=t_1}^{t} \Big( \sum_{F_{d,q}} \lambdai + \sum_{f \in H_{d}^F}  \lambdaa \Big)
\end{equation}

Cost coefficients are calculated in a similar manner.
Rule overhead cost coefficients count the number of installed aggregation rules in the bottlenecked switch. 
They can be set to zero except for $\wTabledZeroOne$ which is set to 1.
Link overhead cost coefficients count the packets matched by the aggregation rule that have to be sent to the remote switch. 
The number of bits processed by a flow rule $f$ in one time slot $t$ is denoted here as $\deltaft$.
Control message overhead cost coefficients count the amount of control messages necessary for delegation.

\begin{equation}
\wLinkdZeroOne := \sumT \sumFdt \delta_{f,t}  \lambdai
\end{equation}

\begin{equation}
\wLinkdOneOne := \sumT \Big( \big( \sumFdt \delta_{f,t}  \lambdai \big) + \big( \sum_{f \in H_{d}^F} \delta_{f,t}  \lambdaa  \big) \Big)
\end{equation}
  
\begin{equation}
\wCtrldZeroOne := 1 + \sumT \sumFdt \lambdai
\end{equation}

\begin{equation}
\wCtrldOneZero :=  1+ \sum_{f \in H_{d}^F} \lambdaa    
\end{equation}
  
\begin{equation}
\wCtrldOneOne :=  \sumT \sumFdt  \lambdai   
\end{equation}

\subsection{Step 2: RS-Alloc}

RS-Alloc takes the output from DT-Select -- which is a set of selected delegation templates $\Dst^*$ for all bottlenecked switches -- and allocates a remote switch for each selected delegation template based on the available free flow table capacity and link bandwidth in the network.
Because this problem is solved globally for all switches once per optimization period, there are multiple  $\Dst^*$-sets that have to be included in the problem formulation.
We use the term $J_t := \bigcup_{s \in S} \Dst^*$ to refer to the total set of all templates. One element $j \in J_t$ is called an allocation job.
And because we are interested in the multi-period problem, we have to extend this definition to a set of time slots $T$.
$T_j \subseteq T$ is a set of consecutive time slot so that allocation job $j$ is active in all time slots in $T_j$. Active means that $J_t$ contains a delegation template with aggregation match $\overrightarrow{m_j}$ that is associated with bottlenecked switch $s_j$.
The first time slot in $T_j$ is always identical to the first time slot in $T$ where a delegation template $d \in D_s$ is selected. And the same delegation template is also selected in all other time slots in $T_j$.
In addition to $T_j$, a second helper construct is needed that collects the different allocation jobs without duplicates. We use $J_T :=  \bigcup_{t \in T} J_t$ for that purpose.

Using this, we define a set of allocation assignments $A_{j,T_j}$ for a single allocation job $j \in J_T$ and a set of consecutive time slots $T_j = \{t_1, \ldots, t_m \}$ as the Cartesian power of the set of all ordered pairs that can be build with $|T_j|$:  $A_{j,T_j}  := R_{j,t_1} \times R_{j,t_2} \times \ldots \times R_{j,t_m}$.
$R_{j,t}$ is the remote set that contains all switches with enough resources in time slot $t$ to handle allocation job $j$. 
Based on $ A_{j,T_j}$, RS-Alloc can be formulated as a multi-dimensional multiple-choice knapsack problem as follows:

\begin{IEEEeqnarray}{rClcccc}
\min_{\Yjrt}  &\quad& \sumJT \sumAjT \wjA ~ Y_{j,A}  &  \label{eq:select-opt:01}\\
\text{s.t.} &	  & \sumAjT Y_{j,A} = 1 & \allJT & \label{eq:select-opt:02} \\
            &     & \uTablert \\
            \IEEEnonumber
            &     &  ~~~+ \sumJt \sum_{A \in \AjT} \alpha_{j\ArrowAllocate r,t} ~ u_{j,t}^{\TTable} ~Y_{j,A} \leq \cTabler & \allT &  ~\allSr &  \label{eq:select-opt:03}  \\
            &     & \uLinkjr \\
            \IEEEnonumber
            &     &  ~~~+ \sumJt \sum_{A \in \AjT} \alpha_{j\ArrowAllocate r,t} ~\dLinkjt ~Y_{j,A}  \leq \cLinkjr & \allT &  ~\allSr  & \label{eq:select-opt:04}  \\
          &     & \uLinkrj \\
          \IEEEnonumber
           &     &  ~~~+ \sumJt \sum_{A \in \AjT} \alpha_{j\ArrowAllocate r,t} ~\dLinkjt ~Y_{j,A}  \leq \cLinkrj \quad\quad & \allT & ~\allSr &  \label{eq:select-opt:05}  \\  
            &     & \Yjrt \in \{0,1\} \quad \allJT ~ \allTj ~ \allSr &  &  &  \label{eq:select-opt:06} 
\end{IEEEeqnarray}

There are $|J_T|$ different mutually disjoint classes $\AjT = \{A_1, \ldots, A_n\}$ called allocation assignment sets. Each allocation job $j \in J_T$ defines one allocation assignment set and each of these sets contains $\prod_{t \in T_j}|R_{j,t}|$ items called allocation assignments. For each allocation job $j \in J_T$ and each time slot $t \in T_j$, there is a utilization coefficient $\dTablejt$. And there is a cost coefficient $\wjA$ for each allocation assignment $A \in \AjT$.
Goal is to choose exactly one item out of each allocation assignment set so that the costs are minimized while the utilization is below the capacity in all time slots. Eq. (\ref{eq:select-opt:01}) ensures that the objective function is minimized. The multiple-choice constraints in Eq. (\ref{eq:select-opt:02}) make sure that exactly one allocation assignment is chosen for each allocation job. Eq. (\ref{eq:select-opt:03}) models the flow table capacity constraints for the remote switches. Eq. (\ref{eq:select-opt:04}) and Eq. (\ref{eq:select-opt:05}) model the link capacity constraints for the links between delegation and remote switches.
All capacity constraints are defined separately for each time slot and each remote switch, i.e., there are $|S|*|T|$ capacity constraints (at most).  And Eq. (\ref{eq:select-opt:06}) ensures that the decision variables are binary.  


%% file: sections/evaluation.tex

\section{Evaluation}
\label{sec:eval}


\subsection{Scenario Generation}
\label{sec:scenario-generation}

We create bottleneck scenarios in two steps. The topology generation step calculates a set of switches $S$, a set of hosts $H$ and the links $Y$ between the switches and hosts based on the Barabasi-Albert model which is often used in the context of software-based networks \cite{eval-ba-use5,eval-ba-use4,eval-ba-use6}. 
This means we use random scale-free topologies where the node degree distribution follows a power law.
The second step creates the flow rules as outlined in Alg. \ref{algo:gen:flow-rules}.

In line \ref{algo:flowrulegen:1}, we first create suitable flow inter-arrival time values $T_\texttt{iat}$ with a gamma distribution based on results presented in \cite{eval-wuerzburg}. 
These values are adjusted in lines \ref{algo:flowrulegen:2} and \ref{algo:flowrulegen:2e} with a global scale parameter $n_\texttt{iat\_scale}$ to adopt it to a multi-switch setting and to add bottlenecks.
$n_\texttt{bneck}$ denotes the number of bottlenecks. For each bottleneck, a random index $x$ is selected between $1$ and $|T_\texttt{iat}|$ as the start of the bottleneck. Next, a second index $y > x$ is selected so that $y$ is minimal and Eq. (\ref{eq:bneck}) is fulfilled, i.e., the bottleneck spans a duration of $n_\texttt{bneck\_duration}$ seconds.

\begin{equation}
 \big\lfloor \frac{\sum_{i = x}^{x+y} T_\texttt{iat}[i]}{1000} \big\rfloor \geq n_\texttt{bneck\_duration}
 \label{eq:bneck}
\end{equation}

\begin{algorithm}[tb!]
    \DontPrintSemicolon
    \SetNoFillComment
    \SetKwRepeat{Assert}{assert}{end}
    \SetKwFunction{FMain}{generate}
    \SetKwProg{Fn}{Function}{:}{end}
    
    \KwData{topology $\big<S,H,Y\big>$ and flow parameters $n_\texttt{xyz}$}
    \KwResult{flow rule set $F_s$ for each switch $s \in S$}
      
    \Fn{\FMain{$\big<S,H,Y\big>, n_\texttt{xyz}$}}{
        $F_s = \{\} ~~ \forall s \in S$\;
        \tcc{Calculate $n_\texttt{pairs}$ inter-arrival time values based on \cite{eval-wuerzburg} and apply scale parameters}
        $T_\texttt{iat} \setTo $ get\_iat\_samples($k$, $\theta$, $n_\texttt{pairs}$)\label{algo:flowrulegen:1}\;
        $T_\texttt{iat} \setTo $ apply($n_\texttt{iat\_scale}$)\label{algo:flowrulegen:2}\;
        $T_\texttt{iat} \setTo $ apply($n_\texttt{bneck},~n_
        \texttt{intensity},~n_\texttt{duration}$)\label{algo:flowrulegen:2e}\;
        $\tau^\texttt{offset} \setTo 10$\label{algo:flowrulegen:tau}\;
        \For{$i \in \{1, \ldots, n_\texttt{pairs}\}$}{\label{algo:flowrulegen:3}
            $\big<\hsrc, \hdst\big> \setTo $ get\_host\_pair($H,~ n_\texttt{isr},~ n_\texttt{hs},~ n_\texttt{hs\_intensity}$)\label{algo:flowrulegen:3a}\;
            
            \tcc{Calculate flow length in bits based on \cite{eval-traffic-19} and select bit rate for the sending host}
            $\delta \setTo $ get\_sample\_from\_jurkiewicz\_mixture()\label{algo:flowrulegen:delta}\;
            $b \setTo $ get\_bitrate($\delta,~n_\texttt{traffic\_scale}$)\label{algo:flowrulegen:b}\;
            
            \tcc{Calculate install and removal time for the new flow rule}
            
            $\tau^\texttt{install} \setTo \tau^\texttt{offset}$\label{algo:flowrulegen:install}\;
            $\tau^\texttt{remove} \setTo \tau^\texttt{offset} + \max\big( \min(\frac{\delta}{b}, 35),~ n_\texttt{lifetime}\big)$\label{algo:flowrulegen:remove}\;
            
            \tcc{Calculate shortest path and add flow rule for each switch on the path}
            $\texttt{sp} = \{s_\texttt{src}, \ldots, s_\texttt{dst}\} \setTo $ get\_sp($\hsrc,~\hdst$)\label{algo:flowrulegen:3d}\;
            \For{$s \in \{s_\texttt{src}, \ldots, s_\texttt{dst}\}$}{
                $f \setTo $new flow rule for $s$\label{algo:flowrulegen:newf}\;
       
                $F_s \setTo F_s \cup \{f\}$\label{algo:flowrulegen:addf}\;
                
            }\label{algo:flowrulegen:3de}
            \tcc{Increase offset by next inter-arrival time value}
            $\tau^\texttt{offset} \setTo \tau^\texttt{offset} + T_\texttt{iat}[i]$\label{algo:flowrulegen:last}\;
        }\label{algo:flowrulegen:3e}
        \Return{$F_s$}
    }
 	\caption{Flow Rule Set Generation}
	\label{algo:gen:flow-rules}
\end{algorithm}

All values in $|T_\texttt{iat}|$ between index $x$ and index $y$ are then reduced to increase the number of flow rules installed in this time period, i.e., they are multiplied with a value $<1$. 
This multiplier is determined by the scale parameter for bottleneck intensity $n_\texttt{bneck\_intensity}$ which is given as a value $>100$. A value of $125$, for example, means that the inter-arrival times in $T_\texttt{iat}$ are multiplied with $\frac{100}{125} = 0.8$ during the bottleneck. Higher intensity values result in lower multipliers and a higher number of installed flow rules per second. To simulate that bottlenecks do not occur instantaneously, the multipliers are modeled as a normal distribution with $\mu = \frac{100}{n_\texttt{bneck\_intensity}}$. Smaller multipliers are used if the indices are close to $x$ and $y$ and larger multipliers are used for indices near $x + \frac{y-x}{2}$.

Lines \ref{algo:flowrulegen:3}-\ref{algo:flowrulegen:3e} then create the individual flow rules, roughly following the methodology used in \cite{eval-infocom-paper}.
Line \ref{algo:flowrulegen:3a} selects a random pair of hosts controlled by the inter switch ratio $n_\texttt{isr}$ which determines the probability that two selected hosts are attached to different switches. Higher inter switch ratios are more likely to cause bottlenecks in switches with high node degree.
We further define a number of hotspots $n_\texttt{hs}$. A hotspot is a subset $H_\texttt{hs} \subseteq H$ and hosts in this subset are preferred in the host pair selection process. For each hotspot, the hosts attached to a randomly selected switch are added to $H_\texttt{hs}$. Each time a new host pair is selected, it is checked whether $\hsrc$ is present in $H_\texttt{hs}$. If this is the case, the selection process is finished. If this is not the case, however, a new pair is selected. The maximum number of re-selections is defined by $n_\texttt{hs\_intensity}$.
Using this mechanism, bottlenecks can be simulated for arbitrary switches -- including those with a low node degree.

Lines \ref{algo:flowrulegen:delta}-\ref{algo:flowrulegen:b} generate the traffic. The bitrate calculation is based on  \cite{eval-traffic-19} and the concrete mixture of uniform and log-normal distributions used to create a sample for $\delta$ in line \ref{algo:flowrulegen:delta} can be found in https://github.com/kit-tm/fdeval/blob/master/topo/static.py\footnote{The original version of this mixture given as a set of json files created by Piotr Jurkiewicz can be found here: https://github.com/piotrjurkiewicz/flow-models/tree/master/data/agh\_2015/mixtures/all/length}.
Given the number of bits $\delta$, the bitrate of the sending host is modeled as a simple square root function.
$n_\texttt{traffic\_scale}$ is used for scenarios where the flow rules process more traffic, e.g., because multiple hosts are aggregated behind one rule. If used, $\delta$ is multiplied with $\frac{100}{n_\texttt{traffic\_scale}}$, i.e., the number of bits is increased linearly.

Lines \ref{algo:flowrulegen:install}-\ref{algo:flowrulegen:remove} finally calculate the install and removal times for the new flow rule based on a global offset variable $\tau^\texttt{offset}$ which is updated in each iteration. Parameter $n_\texttt{lifetime}$ represents the minimum lifetime of all flow rules and is used here if $\frac{\delta}{b}$ is too small. This is similar to the idle timeout mechanism used in OpenFlow networks. Because scenarios are currently limited to 400 seconds, a static maximum lifetime of $35$ seconds is used. 
The procedure continues until there are $n_\texttt{pairs}$ iterations in total.

\subsection{Example Experiment}

\begin{figure}[!htb]
\begin{center}
\includegraphics[width=1\columnwidth]{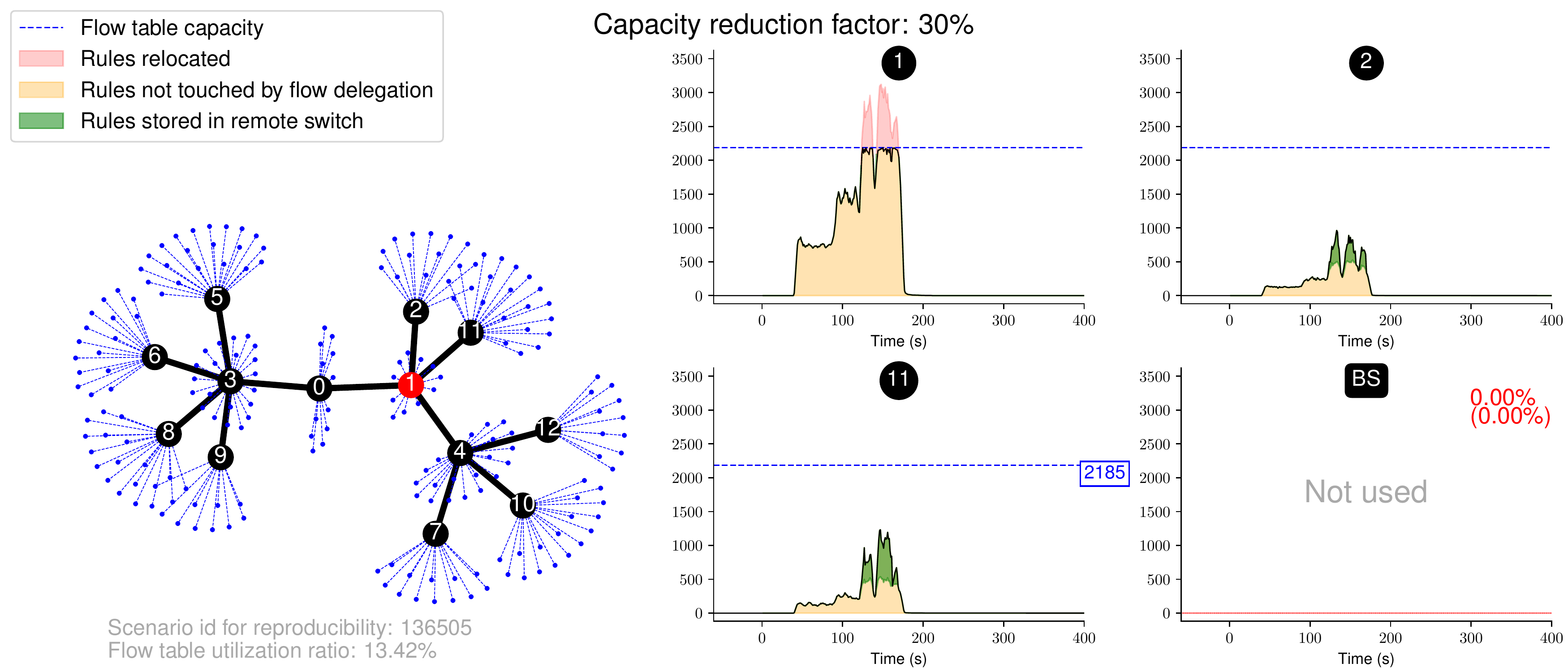}
\caption[caption]{Scenario $z_{136505}$ with capacity reduction of 30\%}
\label{fig:scenario-reduction-136505-30}
\end{center}
\end{figure}

Fig. \ref{fig:scenario-reduction-136505-30} shows an experiment with a maximum flow table utilization of 3122 rules (peak at switch 1) and a maximum flow table capacity of 2185 flow rules shown here as the blue line.
It depicts the situation after the flow delegation algorithm was executed. The first three plots correspond to the topology on the left. The fourth switch labeled BS is a virtual backup switch that is used when the algorithm can not find a suitable remote switch.
The area colored in red in switch 1 denotes relocated rules that were delegated away from the bottlenecked switch. The green area in switches 2 and 11 denote relocated remote rules. The area colored in yellow shows flow rules that are not touched by the algorithm, including additional rules such as aggregation and backflow rules. There are several important observations to make here:
\begin{enumerate}
 \item None of the three switches suffers from a capacity bottleneck when our algorithm is used.
 \item The utilization of the bottlenecked switch (1) is right below the capacity threshold.
 \item The backup switch is not used, i.e., all flow rules are handled successfully.
 \item There is still some free capacity available so we could handle more severe bottlenecks.
\end{enumerate}

The three first observations demonstrate the feasibility of our algorithm for this example. The last observation motivates a new metric to measure bottleneck severity. We introduce the term capacity reduction for this purpose which is explained in more detail in the next section.

\subsection{Capacity Reduction}

Capacity reduction is a percentage value that specifies the relation between maximum required amount of rules and actual flow table capacity for a specific scenario. And for a multi-switch scenario we simply take the maximum capacity reduction of all switches. For the example in Fig. \ref{fig:scenario-reduction-136505-30}, we got a maximum utilization of 3122 rules. And we have a maximum capacity of 2185 rules which is 30\% below the maximum utilization. We say that flow delegation has to cope with a capacity reduction of 30\% in this scenario. Consequently, higher capacity reduction is equivalent to more severe bottlenecks.

\begin{figure}[!htb]
\begin{center}
\includegraphics[width=1\columnwidth]{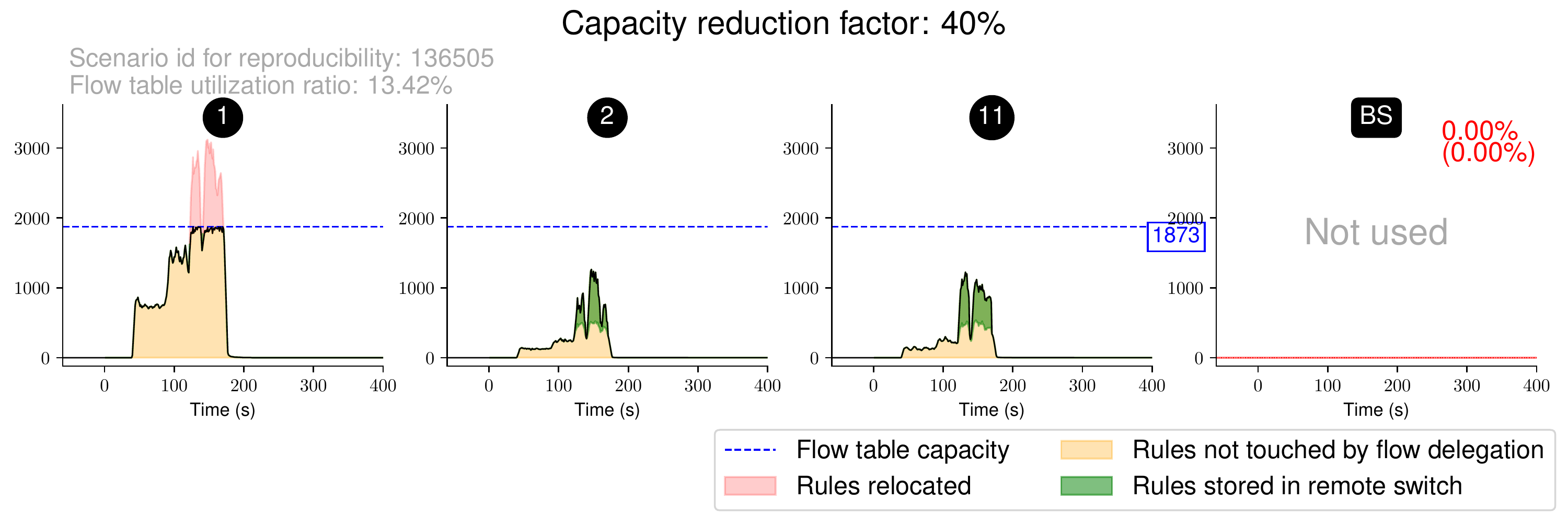}
\caption[caption]{Scenario $z_{136505}$ with capacity reduction of 40\%}
\label{fig:scenario-reduction-136505-40}
\end{center}
\end{figure}

Fig. \ref{fig:scenario-reduction-136505-40} shows the same scenario as before but with a capacity reduction of 40\% instead of 30\% which is achieved by lowering the capacity from  2185 to 1873. It can be seen that the same two remote switches are chosen. And while the backup switch is still at 0\%, the flow table utilization of remote switches 2 and  11 is higher than in Fig. \ref{fig:scenario-reduction-136505-30} because of the increased number of rules relocated from switch 1.

\subsection{Failure Rate}

It is quite obvious from the last two examples that more and more severe bottlenecks will eventually lead to a situation where the flow delegation algorithm is not able to find remote switches with enough spare capacity. 
To measure this aspect, the so-called failure rate counts the amount of flow rules that can not be assigned to a remote switch. This value is multiplied with a  normalization factor that represents all used flow rules over the course of the experiment to get a percentage value. A failure rate of $0.1\%$, for example, means that flow delegation was not able to find a suitable remote switch for $0.1\%$ of the flow rules.

\begin{figure}[!htb]
\begin{center}
\includegraphics[width=1\columnwidth]{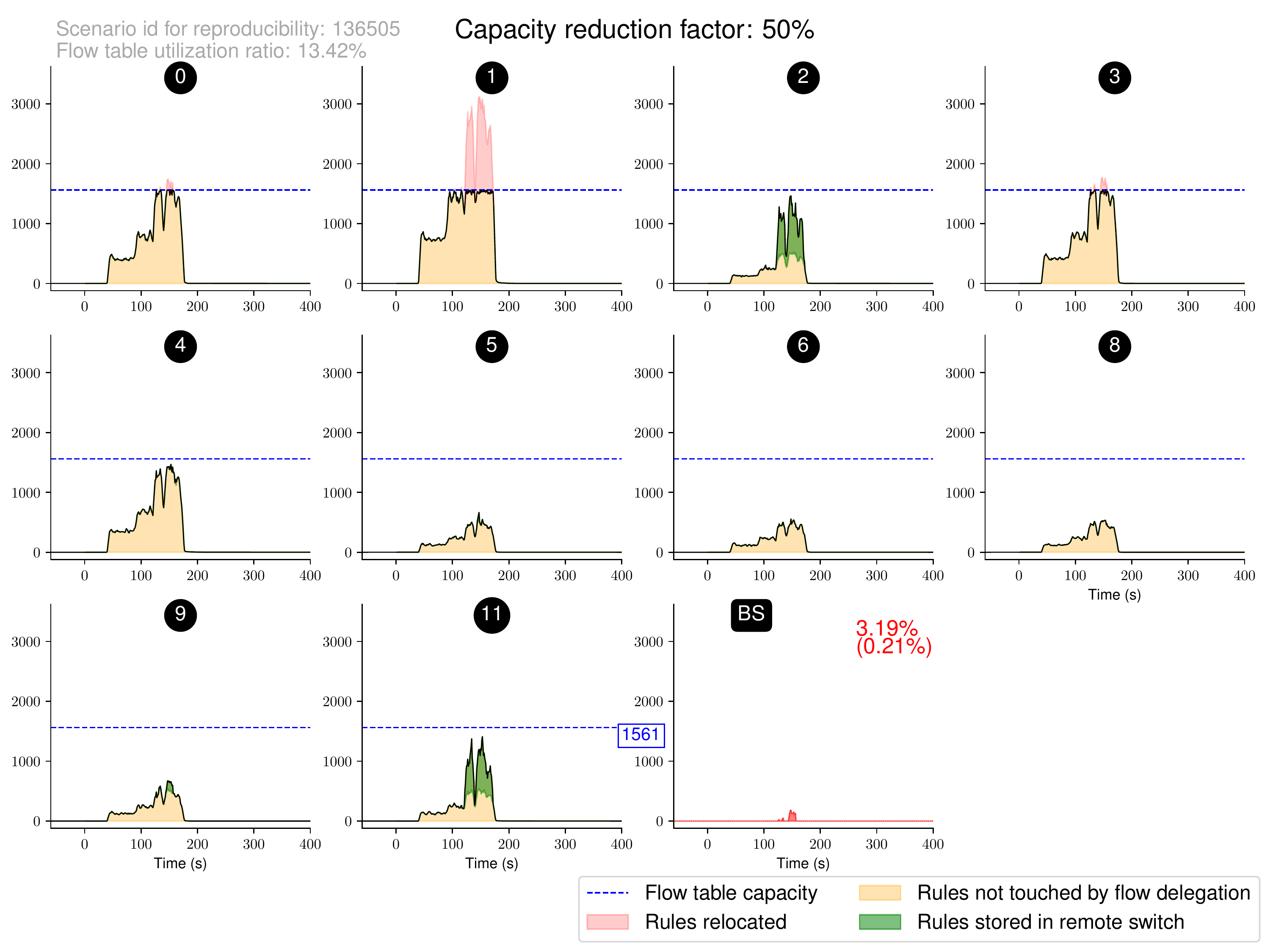}
\caption[caption]{Scenario $z_{136505}$ with capacity reduction of 50\%}
\label{fig:scenario-reduction-136505-50}
\end{center}
\end{figure}

Fig. \ref{fig:scenario-reduction-136505-50} shows the example scenario with a capacity reduction factor of 50\%. In this case, 10 of the 13 switches interact with the flow delegation algorithm. 
Unlike before, there are now multiple bottlenecked switches (0, 1, and 3).
The previously used remote switches (2, 11) are highly utilized during the bottleneck phase and there is not much free flow table capacity left. In fact, all neighboring switches of switch 1 are highly utilized.
Several other switches such as 5, 6, and, 8 still provide spare resources. They can, however, only be used for the switch 3 bottleneck due to the physical connections in the topology.
Not all flow rules can be successfully relocated to a remote switch any more. The failure rate in this experiment is $0.21\%$, i.e., $0.21\%$ of all flow rules -- or 3.19\% of the to be relocated flow rules in the red area -- are relocated to the backup switch.

\subsection{Performance}

In this and the following sections, we evaluate how our algorithm performs for 4996 different bottleneck scenarios.
All scenarios were created according to Sec. \ref{sec:scenario-generation} and are publicly available at https://publikationen.bibliothek.kit.edu/1000120288. 
We used dataset D5000 for our evalution. The code, all parameters and detailed instructions for reproducibility can be found at https://github.com/kit-tm/fdeval.

To evaluate raw performance, we compare our approach to a baseline algorithm that uses the optimization problem defined in Eq. (\ref{eq:dssel:ilp:obj}) - (\ref{eq:dssel:ilp:binary}) as step 1, i.e., DT-Select always computes the optimal results for the current optimization period. Note that this is not applicable in practice due to the large computational effort that is required to solve this problem (up to several hours for a single scenario). We also compare the algorithm to a greedy strategy that mimics the behavior of the threshold-based delegation algorithm described in \cite{own-pbce}.

\begin{figure}[!htb]
\begin{center}
\includegraphics[width=1\columnwidth]{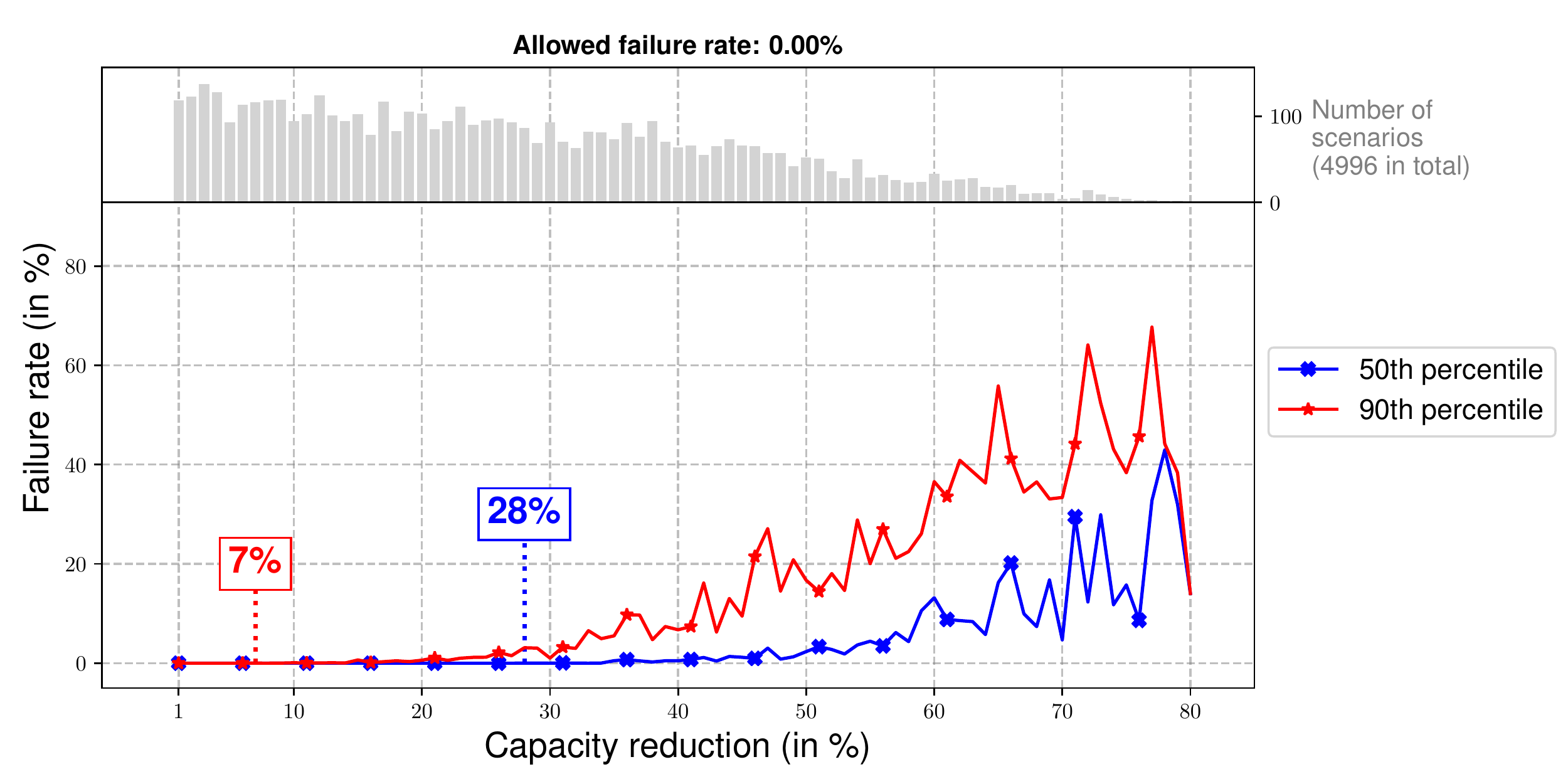}
\caption[caption]{Capacity reduction and failure rates for 4996 different scenarios}
\label{fig:ds_5000_class_0_100_0}
\end{center}
\end{figure}

The setup is as follows. We first take the 4996 scenarios and order them by capacity reduction. Recall that we use this value to measure bottleneck severity.
 This results in 80 groups between 1\% (small bottlenecks) and 80\% capacity reduction (large bottlenecks). 
These groups are shown on the x-axis in Fig. \ref{fig:ds_5000_class_0_100_0}. 
The grey histogramm in the top denotes the amount of scenarios contained in each group. 
There is an (intended) bias towards smaller capacity reduction values in the used dataset because values above 50\% are not our main focus. For very high values above 70\%, most of the groups contain less than 5 entries and are thus not considered representative.
Next, we calculate the failure rate for each scenario in each group.
We then plot the 50th and 90th percentile of the failure rates for all groups. The results are shown on the y-axis in Fig. \ref{fig:ds_5000_class_0_100_0}. 
The colored rectangles denote the highest group that could be handled with a 0\% failure rate.
It can be concluded from the blue rectangle that flow delegation achieves  a failure rate of 0\% for  50\% of the scenarios with capacity reduction of 28\% and lower.
Similarly, the red rectangle indicates that 90\% of the scenarios can be handled with a 0\% failure rate for capacity reduction up to 7\%.

The 7\% value (red) can be seen as worst case baseline. It can be expected that, regardless of the actual situation, our flow delegation algorithm can deal with a bottleneck that falls into the 7\% reduction range while maintaining a 0\% failure rate. This range can be extended significantly when failure rates slightly above 0\% are allowed (15\% for 0.1\% failure rate and 21\% for 1\% failure rate) but this is not discussed here in more detail. The 28\% value (blue) is meant as an optimistic expectation that the majority of the bottleneck situations in the 28\% range can be handled without failures.
While 7\% and 28\% does not sound impressive at first glance, the potential gain in practice is significant. Assume an existing network is equipped with switches that support 1000 flow rules.
For this network, 28\% reduction is sufficient to mitigate bottleneck situations with 1387 flow rules, i.e., more than 38\% above the maximum capacity of the switches.
28\% reduction also means the network operator can replace the switches with a smaller (cheaper) model that only supports 720 flow rules, i.e., if the 1000 rules are only required in rare high load scenarios.

\vvv
Another important observation from  Fig. \ref{fig:ds_5000_class_0_100_0} is that higher capacity reduction in the area at and above 50\% lead to unacceptable high failure rates. Flow delegation cannot deal with such extreme situations because there is simply not enough free capacity available. 
It can further be seen that the y-values increase ``slowly'' for x-values between 7\% and approx. 20\% in case of the red curve. This means that only a small fraction of the flow rules can not be allocated to a remote switch. A similar observation can be made for the blue curve up to an x-value of approx. 40\%.

\subsection{Overhead}

\begin{figure}[!htb]
\begin{center}
\includegraphics[width=1\columnwidth]{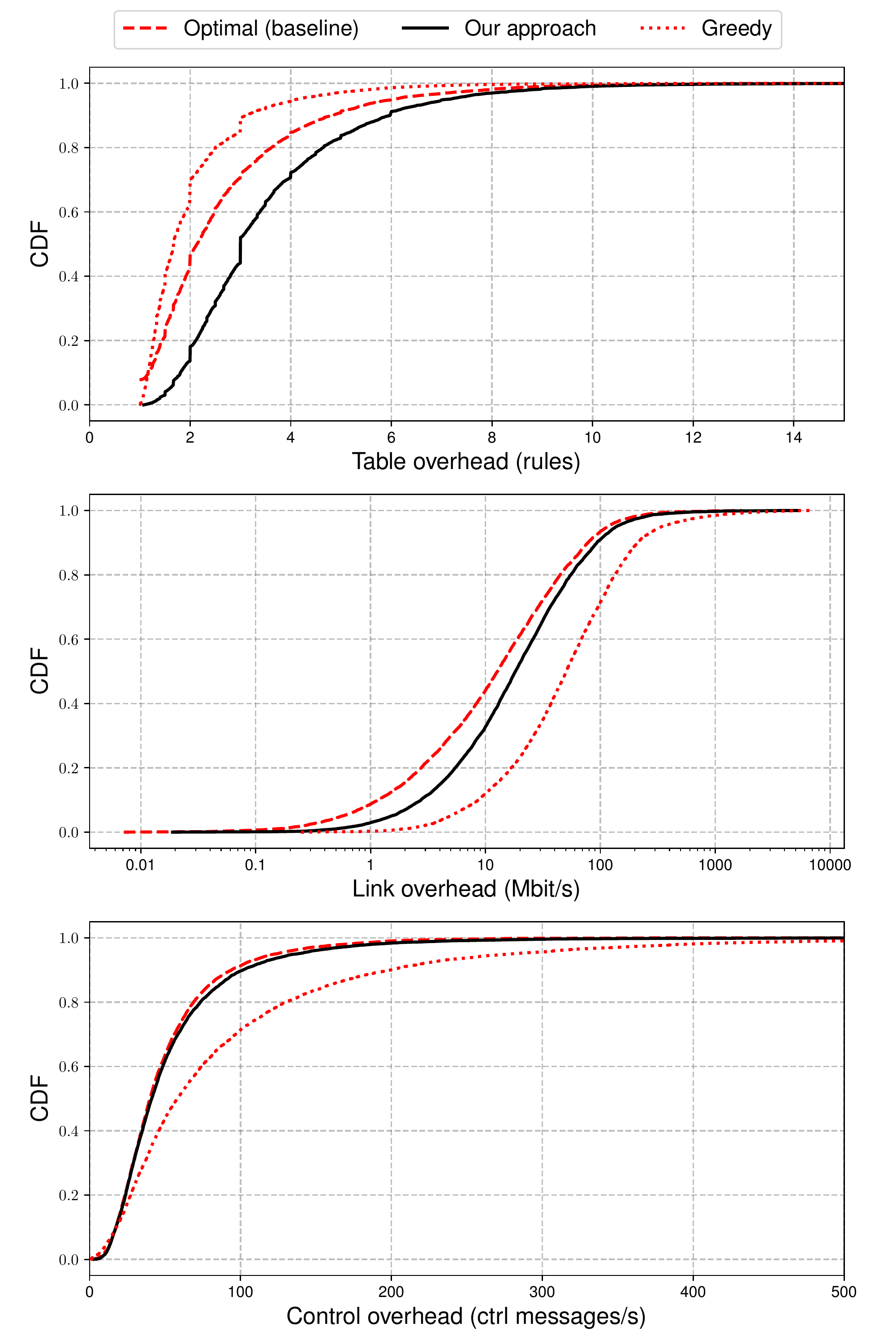}
\caption[caption]{Overhead of the flow delegation algorithm for 4996 different scenarios}
\label{fig:eval:overhead}
\end{center}
\end{figure}

There are three major kinds of overhead considered in our algorithm: the additional rules installed in the flow table of the delegation
switch (table overhead), the traffic relocated from delegation switch to remote
switch and vice versa (link overhead), and the interactions between controller and switches in terms of additional control messages for the delegation process (control overhead). 
The measured overhead for all scenarios is shown in Fig. \ref{fig:eval:overhead}. The x-axis denotes the overhead metric. The y-axis represents the fraction of experiments where this metric is below the x-value.

It can be seen that our flow delegation algorithm (solid black line) performs significantly better than the greedy strategy when it comes to link and control overhead. For control overhead, the results are almost identical to the optimal results.
The fact that the greedy strategy has the lowest table overhead is linked to its lack of flexibility: it can only select or unselect delegation templates in one time slot, i.e., it cannot select one template in exchange for another one. It therefore tends to just stick with a once selected delegation template. This means most delegation templates are used for a long time which again means that more rules of this template are relocated. Because more relocated rules are associated with a single template, less templates are required (lower table overhead). However, at the same time, the two other metrics are significantly worse when such a simple strategy is used.

\subsubsection{Table Overhead}
We can handle 99\% of the investigated scenarios with less than 10 aggregation rules on average. 
In other words: there are approx. ten aggregation rules present in the bottlenecked switch to forward traffic to the remote switch.
For 50\% of the scenarios, the table overhead is 2.13 rules, compared to 3 rules in the optimal case. Consequently, the total amount of additional rules stays below 30 rules for a 20 port switch in almost all cases if the static backflow rules are considered as well. This is less than 1\% of the capacity in case of a 3000 rule flow table. Table overhead is therefore not considered a critical limitation. 

\subsubsection{Link Overhead}
The amount of relocated traffic is below 18.26 Mbit/s per bottlenecked switch on average in 50\% of the investigated scenarios, compared to 12.74 Mbit/s in the optimal case and 50.63 Mbit/s in the greedy case. Note that this additional traffic can be split over multiple links if more than one remote switch is used.
However, the amount of additional traffic depends heavily on the scenario under investigation and can be significant. In 1\% of the scenarios, we measured values above 363.19 Mbit/s.
This happens if no aggregation rules with a low traffic profile are available, i.e., there are too many elephant flows. 
In such extreme scenarios, flow delegation might not be applicable if link bandwidth is a critical concern.


\subsubsection{Control Overhead}
The amount of additional control messages per time slot is below 42 in 50\% of the investigated scenarios, compared to 40 in the optimal case (below 74 in 80\% of the scenarios and below 237 in 99\% of the scenarios). Modern switches can deal with this kind of overhead.
Note that new flow rules that should be installed in the delegation switch while delegation is enabled are also counted as ``additional'' flow rules because the rule is redirected to another switch. 

\subsection{Runtime}

For the runtime evaluation, we analyze solving and modeling times for both, DT-Select and RS-Alloc. 
Solving time is the time spent inside the ILP solver which is Gurobi in our case. 
Modeling time is the time required to pre-process the input data, prepare the model and post-process the output of the solver.
The reported values refer to one optimization period or one ``sample''. 
Note that the algorithm is executed periodically and multiple samples must be calculated for each step: n samples for DT-Select where n is the number of the bottlenecked switches and 1 sample for RS-Alloc. 
The experiments are executed on a server with 2 x Intel(R) Xeon(R) Silver 4110 CPUs (32 logical cores, clocked at 2.10 GHz), restricted to a single CPU core.

\begin{figure}[!htb]
\begin{center}
\includegraphics[width=1\columnwidth]{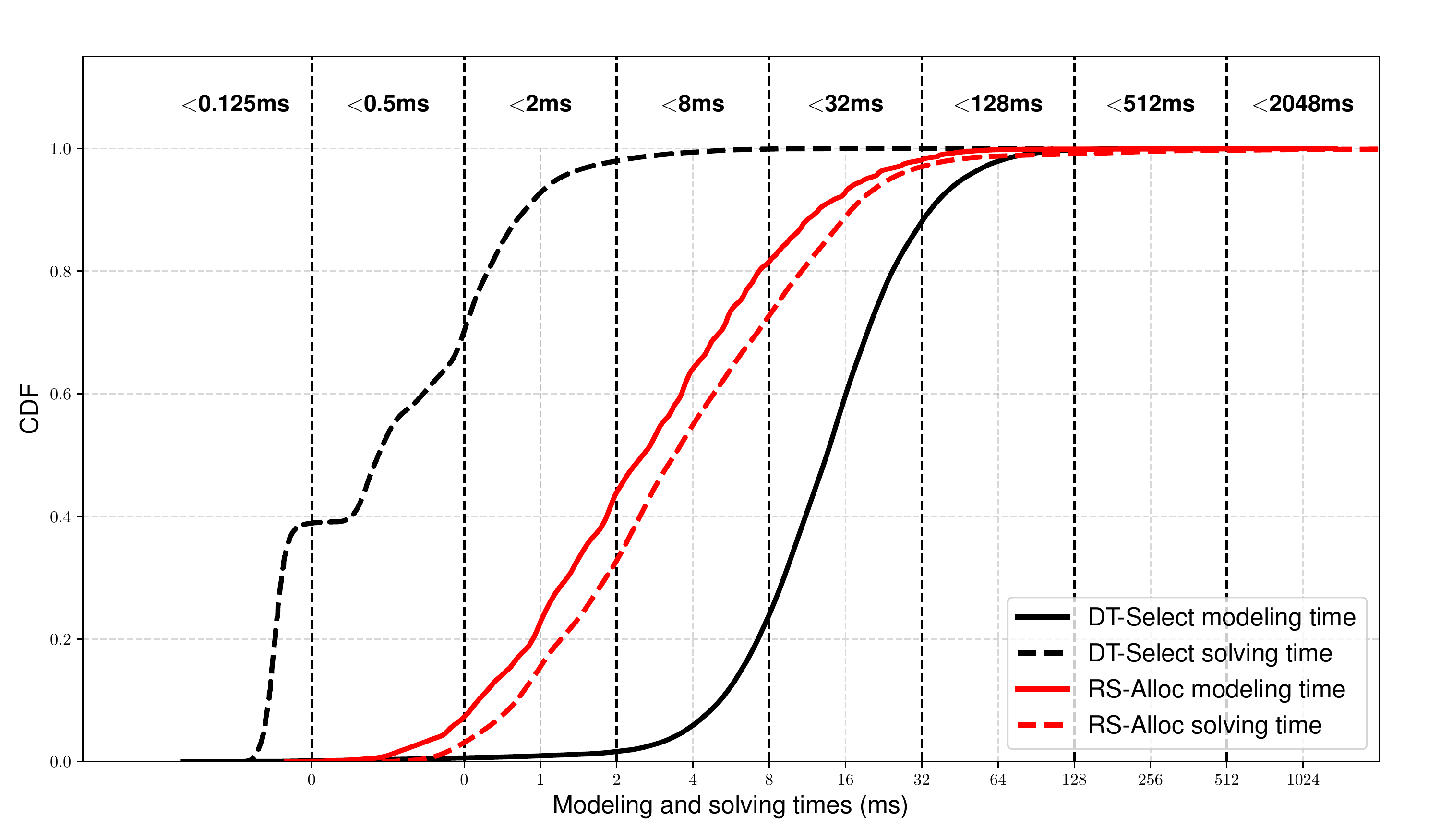}
\caption[caption]{Measured runtime for 4996 scenarios with 1176504 samples of DT-Select and 364029 samples of RS-Alloc}
\label{fig:eval:runtime:dts-modeling}
\end{center}
\end{figure}

Fig. \ref{fig:eval:runtime:dts-modeling} shows the measured modeling and solving times.
For the investigated 4996 scenarios, there are 1.176.504 samples of DT-Select and 364.029 samples for RS-Alloc. 
The x-axis denotes measured time in milliseconds on a logarithmic scale. The y-axis denotes the fraction of the samples with a modeling / solving time below the specified x-value. 

It can be observed that the vast majority of the samples is computed in less than 128ms using a single CPU core. This is fast enough to run the algorithm once per second. And because the DT-Select samples are independent from each other, they can be executed in parallel on multiple CPU cores. It can therefore be concluded that our approach is fast enough for practical application.
It can further be seen that the runtime is dominated by the DT-Select modeling step (solid black curve). 
99.78\% of the DT-Select samples have a modeling time below 128ms. This means only 0.22\% of the samples require more than 128ms with a worst case modeling time across all samples of 383ms. 
However, these number can be optimized in several ways in a real system. For once, the current modeling is done completely in python without optimizations and contains logging/debugging code that could be removed. Using a just-in-time compiler such as PyPy or a more efficient programming language will probably also reduce the modeling time significantly. However, this was not tested. Instead, it was investigated which conceptual part of the modeling step contributes the most to the final measurement result. 
This investigation showed that more than 50\% of the modeling time in the current prototype is used for creating the link overhead cost coefficients. This is expensive because data for each individual flow rule object is extracted. In a real system, however, this data can be provided by some monitoring component and it is not required to calculate the coefficient from scratch in each optimization period. This and other implementation specific aspects provide potential for optimization.


%% file: sections/related_work.tex

\section{Related Work}
\label{sec:rw}

The work closest to ours are caching algorithms for software offloading which is a heavily studied research area in the context of SDN. Software offloading is based on the observation that fast hardware and flexible software can complement each other.
Flow rules processing a large amount of packets are stored in the hardware flow table. The remaining flow rules are offloaded to a software flow table.  
The goal of the caching algorithms is it to select flow rules that should be offloaded to the slower software table while existing dependencies are not violated. This is very similar to what the DT-Select step has to perform in our algorithm.

A well-known approach in this category is CacheFlow \cite{KARW14-cacheflow}.
The CacheFlow architecture uses one or multiple software switches attached to a hardware switch. The latter is then used as a cache for the most popular flow rules in the network. The caching algorithm is responsible to decide which flow rules are placed in the cache and which are placed in the slower software switches. 
CoverSet \cite{rule-caching-algos-14} uses newly created dummy flow rules to reduce the number of dependencies prior to the caching decision. 
The authors in \cite{ShCh16-rule-cache-replacement} extend this approach so it can work on a set of flow rules instead of only individual flow rules. In addition, their algorithm exploits temporal and spatial localities. Temporal locality means a flow rule that was triggered by network traffic will be triggered again soon and spatial locality means that, at short time scales, traffic concentrates on flow rules with similar packet header fields.
CAB \cite{cab-14, cab-18} chooses a different approach based on a geometric representation of the set of flow rules -- hyper-rectangles where each dimension  refers to one packet header field -- to deal with dependencies. 
Other researchers propose two-stage caching architecture that rely on multiple smaller TCAM chips in the hardware flow table to further optimize caching \cite{craft-17, balancer-17, nvtcam-17} . 
And there is a range of other approaches that go in a similar direction \cite{fdrc-15, cost-minimization-rule-caching-16, cnor-18, cuca-19}.

%


Software offloading and the associated caching algorithms can theoretically increase flow table capacity by orders of magnitude which is not possible with our flow delegation algorithm.  The maximum number of extra flow rules that can be provided by any flow delegation algorithm depends on the available free capacity in the network. Say the bottleneck switch has a flow table capacity of $N$. We may be able to deal with a demand of $2*N$ or $3*N$, but not with $100*N$. 
However, software offloading has two severe drawbacks: i) additional infrastructure in form of software tables has to be installed in the network and ii) flow rules placed in a slower software flow table suffer from a significant performance degradation~\cite{study_latency, pisces, study_ovs}. This is especially critical if the software flow table is not placed in immediate vicinity of the hardware flow table which can result in unacceptable delay. 
In comparison, our approach requires no infrastructural changes. Performance impact on delegated flows is small (up to 0.1$ms$ of additional delay) and insusceptible to fluctuations because processing of the packets is still done in hardware.
Furthermore, flow delegation can be easily combined with software offloading by using a software table as remote switch.

Other related work besides caching algorithms and software offloading mainly falls in three different categories which are briefly outlined below.

\subsection{Flow Rule Eviction}
\label{sec:background:rw3}

The idea behind flow rule eviction algorithms is it to remove an existing and potentially still active flow rule from the flow table if a new flow rule has to be installed and there is no free flow table capacity available. 
This is usually done by calculating proper timeouts for each flow rule.
SmartTime \cite{smarttime-14} uses a heuristic to determine TCAM-efficient idle timeouts. The idea is to proactively evict flow rules from the flow table before a bottleneck occurs. 
TimeoutX \cite{timeoutx-15} calculates adaptive timeouts for new flow rules based on estimated flow rule lifetime and current flow table utilization. 
The authors in \cite{ml-eviction-18} and \cite{ml-eviction-19} use machine learning to tune timeouts for proactive flow rule eviction. 
And there are many other algorithms that manipulate timeouts to prevent bottlenecks \cite{ahtm-14, intelligent-timeout-master-15, timeout-liu-16, timeout-xu-18}.
The authors in \cite{eviction-bloomfilter-16} utilize bloom filters stored in SRAM to track the importance of existing flow rules. FlowMaster \cite{flowmaster-14} proposes a per flow prediction algorithm to assess the importance that can be implemented in TCAM/SRAM. IRCR \cite{ircr-18} uses a new software module at the SDN switch that considers the arrival probability of new traffic for existing flow rules to determine the best eviction candidate.

All these approaches have the problem that the available flow table capacity of a bottlenecked switch is not actually increased. Eviction will just remove ``less important`` flow rules from the flow table -- which only help with scalability if such less important flow rules exist. It cannot deal with situations where a switch is highly utilized and all flow rules are actively required. Even worse, installing and evicting a large number of flow rules in short succession can lead to a phenomenon called  ``rule replacement storm'' \cite{flow-aware-routing-18} which has significant negative impact on scalability. 

\subsection{Flow Rule Distribution}
\label{sec:background:rw4}

These approaches try to distribute the set of required flow rules among all available hardware switches.
The most notable work in this area is DIFANE \cite{YRFW10-difane} which achieves scalability by distributing flow rules over a set of authority switches. The controller first distributes the space of all possible flow rules (flowspace) into several partitions and each authority switch is responsible for only one partition. All flow rules are then proactively installed at their respective authority switch.
Palette \cite{KaHK13-palette} follows a very similar approach and distributes large flow tables into equivalent sub-tables that are then distributed among the hardware switches.
Similar ideas can also be found in \cite{KLRW13-one-big-switch, optimize-rule-placement-14, LHWX+15-ftrs, rule-distribution-16, rule-distribution-18, rule-distribution-sat-18}.

This is similar to our algorithm in the sense that it tries to efficiently utilize existing flow table capacity in the network. However, flow rule distribution systems usually performs a ``re-design'' of the network from a control plane perspective so that existing applications are restricted to the specific design of the solution. As a result, none of the above approaches can be used with arbitrary network applications. Most of them also require infrastructural changes or changes to the southbound interface.

\subsection{Flow Table Compression}
\label{sec:background:rw5}

Approaches based on compression try to minimize the number of required flow rules without violating the high level decisions from the network applications. 
Unlike flow rule distribution, compression usually focuses on a single switch / flow table.
TCAM Razor \cite{LiMT10-tcam-razor} takes a set of input flow rules and uses a multi-step process based on decision diagrams, dynamic programming and redundancy removal to create a smaller but semantically equivalent set of output flow rules. 
Bit Weaving \cite{MeLT12-TCAM-compaction-bit-weaving} is based on the idea that adjacent flow rules with the same action and a hamming distance of 1 and can be merged into a single flow rule. 
Compact TCAM \cite{KaBa13-compact-tcam} reduces the number of bits occupied by a single flow rule in the hardware flow table. It is based on the idea that flows in the network are classified based on static-sized, unique flow identifiers instead of arbitrary packet header fields (which can require hundreds of bits). 
\cite{wildcard-compression-14} relies on wildcard aggregation based on the Espresso heuristic which is usually used for logic minimization. 
Minnie \cite{minnie-15} uses wildcard rules to compress existing routing tables based on source and destination addresses. 
And other work focusses on generic TCAM compression independently of SDN \cite{packet-classifier-smaller-06, tcam-compaction-ruleset-min-09, tcam-compaction-gray-coding-12}.

Flow table compression can reduce the number of required flow rules but their efficiency depends heavily on the use case.
Generic compression schemes such as TCAM Razor report average savings of 81.8\% \cite{LiMT10-tcam-razor}. This is possible because the evaluation focuses on range expansion which is not that relevant for SDN. Compression schemes that focus on SDN report savings between 17\% \cite{wildcard-compression-14} and 50\% \cite{minnie-15}. Compression-based approaches, however, suffer from the same problem as flow rule distribution: the compression process will change the set of flow rules which is  not transparent to the control plane. This hinders compatibility and the approach cannot be used with arbitrary network applications.

%% file: sections/conclusion.tex
\section{Conclusion}
\label{sec:conclusion}

This paper introduces a new delegation template abstraction for flow delegation that is much easier to handle than the output of comparable approaches such as dependency graph algorithms \cite{rule-caching-algos-14}. Such delegation templates are not only useful for flow delegation but could also be applied in the context of software offloading  \cite{BoPa12-autoslice, KARW14-cacheflow, memory-swapping-17}, flow rule distribution \cite{YRFW10-difane, KaHK13-palette} or other related areas.
The paper also introduces an efficient and fast algorithm for flow delegation based on the new abstraction that deals with multiple and potentially conflicting objectives and can consider future network situations to proactively mitigate anticipated bottlenecks. 
We evaluate 4996 different bottleneck scenarios with the new algorithm and show that it performs significantly better than previous approaches that used a simple greedy heuristic. We also show that the algorithm is fast enough to be used in practice: a single commodity server with 32 cores can handle a network with hundreds of switches.